\documentclass[aps,prd,reprint,superscriptaddress,nofootinbib]{revtex4-2}
\usepackage{amsmath,amsfonts,amssymb,bm}
\usepackage{graphicx,subfigure}
\usepackage{bbold}
\usepackage[normalem]{ulem}
\usepackage[dvipsnames,usenames]{xcolor}

\usepackage[colorlinks,citecolor=blue]{hyperref}

\begin{document}

\title{Evolution of collisional neutrino flavor instabilities in spherically symmetric supernova models}

\author{Zewei Xiong}
\email[Email: ]{z.xiong@gsi.de}
\affiliation{GSI Helmholtzzentrum {f\"ur} Schwerioneneforschung, Planckstra{\ss}e 1, 64291 Darmstadt, Germany}

\author{Meng-Ru Wu}
\affiliation{Institute of Physics, Academia Sinica, Taipei 11529, Taiwan}
\affiliation{Institute of Astronomy and Astrophysics, Academia Sinica, Taipei 10617, Taiwan}
\affiliation{Physics Division, National Center for Theoretical Sciences, Taipei 10617, Taiwan}

\author{Gabriel Mart{\'i}nez-Pinedo}
\affiliation{GSI Helmholtzzentrum {f\"ur} Schwerioneneforschung, Planckstra{\ss}e 1, 64291 Darmstadt, Germany}
\affiliation{Institut f{\"u}r Kernphysik (Theoriezentrum), Fachbereich Physik, Technische Universit{\"a}t Darmstadt, Schlossgartenstra{\ss}e 2, 64289 Darmstadt, Germany}

\author{Tobias Fischer}
\affiliation{Institute of Theoretical Physics, University of Wroclaw, Pl. M. Borna 9, 50-204 Wroclaw, Poland}

\author{Manu George}
\affiliation{Institute of Physics, Academia Sinica, Taipei 11529, Taiwan}

\author{Chun-Yu Lin}
\affiliation{National Center for High-performance Computing, National Applied Research Laboratories, Hsinchu Science Park, Hsinchu City
30076, Taiwan}

\author{Lucas Johns}
\affiliation{Departments of Astronomy and Physics, University of California, Berkeley, CA 94720, USA}

\date{\today}

\begin{abstract}
We implement a multigroup and discrete-ordinate neutrino transport model in spherical symmetry which allows to simulate collective neutrino oscillations by including realistic collisional rates in a self-consistent way.
We utilize this innovative model, based on strategic parameter rescaling, to study a recently proposed collisional flavor instability caused by the asymmetry of emission and absorption rates between $\nu_e$ and $\bar\nu_e$ for four different static backgrounds taken from different stages in a core-collapse supernova simulation.
Our results confirm that collisional instabilities generally exist around the neutrinosphere during the supernova accretion and postaccretion phase, as suggested by Johns [arXiv:2104.11369.]. 
However, the growth and transport of flavor instabilities can only be fully captured by models with global simulations as done in this work.
With minimal ingredient to trigger collisional instabilities, we find that the flavor oscillations and transport mainly affect (anti)neutrinos of heavy lepton flavors around their decoupling sphere, which then leave imprints on their energy spectra in the free-streaming regime.
For electron (anti)neutrinos, their properties remain nearly intact.
We also explore various effects due to the decoherence from neutrino-nucleon scattering, artificially enhanced decoherence from emission and absorption, neutrino vacuum mixing, and inhomogeneous matter profile, and discuss the implication of our work.
\end{abstract}

\maketitle
\graphicspath{{./figures/}}

\section{Introduction}
\label{sec:introduction}
The phenomenology of neutrino flavor oscillations has been established by experiments with solar, atmospheric, reactor, and accelerator neutrinos~\cite{pdg}.
From these experiments, it is well understood that flavor oscillations depend not only on the properties of neutrino mixing in vacuum, but also the coherent forward scattering of neutrinos with electrons in medium, e.g., the earth mantle or the solar interior.

In astrophysical environments such as core-collapse supernovae (CCSNe) and binary neutron star mergers (BNSMs), neutrino fluxes are sufficiently intense so that forward scattering among neutrinos is important and cannot be ignored.
The nonlinear interaction among neutrinos themselves leads to various collective phenomena and flavor instabilities in two main categories: the so-called ``slow'' mode (e.g., Refs.~\cite{pastor2002physics,Duan:2005cp,Hannestad:2006nj,Duan:2006an,duan2007picture,raffelt2007self,Dasgupta:2007ws,Gava:2009pj,Friedland:2010sc,Raffelt:2013rqa,Duan:2013kba,Wu:2014kaa,Mirizzi:2015fva,Abbar:2015fwa,Dasgupta:2015iia,Martin:2019kgi}), and ``fast'' flavor conversion (e.g., Refs.~\cite{sawyer2005speed,Sawyer:2015dsa,Dasgupta:2016dbv,Chakraborty:2016lct,Izaguirre:2016gsx,Capozzi:2017gqd,abbar2018fast,Airen:2018nvp,johns2020neutrino,xiong2021stationary,bhattacharyya2021fast}); see also review papers~\cite{duan2010collective, chakraborty2016collective,tamborra2020new,Capozzi:2022slf,Richers:2022zug} and references therein.
Both categories require a crossing from positive to negative values in the distribution of the neutrino electron lepton numbers ($\nu$ELN)~\cite{Morinaga:2021vmc,Dasgupta:2021gfs}.
The distinction is that the slow mode relies on the crossing in energy spectrum while the fast mode requires one in the angular distribution. 
For the slow mode, it typically has a flavor conversion length scale of $\mathcal{O}(10-100)$~km and occurs far outside the neutrinosphere after neutrinos decouple from the medium. 
For the fast mode, it converts neutrino flavors in a much shorter length scale $\sim (G_F n_\nu)^{-1} \sim \mathcal{O}(1)$~cm with $G_F$ the Fermi constant and $n_\nu$ the neutrino number density.
Studies have shown that fast instabilities generally exist in certain regions near or even inside the neutrinosphere in CCSNe modeled by multidimensional simulations~\cite{abbar2019occurrence,azari2020fast,morinaga2020fast,Abbar:2019zoq,Glas:2019ijo,nagakura2021occurrence,harada2022prospects}, 
and are even more ubiquitously present in the postmerger environments of BNSMs~\cite{Wu:2017qpc,wu2017imprints,George:2020veu,Li:2021vqj,Just:2022flt,Richers:2022dqa,Fernandez:2022yyv,grohs2022neutrino}.
The potential importance and the associated very short length scale of fast mode catalyzes recent surges of local dynamical simulations in tiny boxes \cite{martin2020dynamic, bhattacharyya2021fast, bhattacharyya2020late,Richers:2021nbx,wu2021collective,Richers:2021xtf,zaizen2021nonlinear, abbar2022suppression,Richers:2022bkd,Bhattacharyya:2022eed,grohs2022neutrino}.

The change of neutrino flavor content due to oscillations can affect the matter composition and the energy exchange of neutrinos with medium through collisional processes, which can also lead to feedback effects on the flavor evolution of the neutrino gas.
A self-consistent treatment including both coherent flavor evolution and collisions have been formulated as the neutrino quantum kinetic equation (QKE) \cite{sigl1993general,Vlasenko:2013fja,Volpe:2015rla,blaschke2016neutrino}. 
Studies that solve simplified QKE by including reduced set of collisions have been done in recent years~\cite{Capozzi:2018clo,Richers:2019grc,martin2021fast, shalgar2021change, Sigl:2021tmj, kato2022effects, sasaki2022detailed}.
Among those, Ref.~\cite{martin2021fast, shalgar2021change, kato2022effects, sasaki2022detailed} examined the impact of neutrino-nucleon scattering (NNS) on the evolution of fast flavor oscillations under the condition that fast instabilities exist a priori.

Recently, a novel kind of flavor instability caused by the asymmetry of emission and absorption (EA) rates between electron neutrinos $\nu_e$ and electron antineutrinos $\bar\nu_e$ has been proposed~\cite{johns2021collisional}.
Together with an additional assumption that heavy lepton neutrinos may have a smaller number density than that of $\bar\nu_e$ in regions where neutrinos decouple, it was found that this collisional flavor instability can trigger flavor conversion within a length scale in the order of inverse mean free path from EA processes, i.e., $\sim\mathcal O$(1)~km near the neutrino sphere. 
Although this length scale is much longer than that of the fast mode, the required condition for the collisional instability to occur completely differs from that of fast instabilities such that both mechanisms can work independently. 
In particular, it can potentially trigger flavor conversions in regions with low electron number fraction ($Y_e$) where neutrinos decouple in CCSNe~\cite{johns2021collisional}.

The pioneer works in Refs.~\cite{johns2021collisional,johns2022collisional} on the collisional instability were based on the assumption of spatial homogeneity.
One general question, however, can be raised: How can a simulation including collisions near the neutrino sphere be performed self-consistently without including the global advection of neutrinos?
This question is not restricted to the collisional instability but applicable to all long-term simulations including both collisions and anisotropic angular $\nu$ELN distribution.
The fundamental reasons that trigger the flavor instabilities such as the diluted heavy lepton neutrino number flux due to diffusion or the anisotropic angular distributions are consequences of neutrino advection in the presence of collisional neutrino processes and inhomogeneity in the length scale of astrophysical environment~\cite{johns2022self}. 
Thus, when the advection of neutrinos is neglected, those conditions for flavor instabilities may not be maintained self-consistently such that the outcome of the long-term neutrino flavor evolution simulations can be inaccurate.
To address this issue, several simulations that aimed to examine flavor oscillations triggered by fast and/or slow modes with the inclusion of global advection in spherically symmetric geometry under given hydrostatic radial profiles have been reported lately~\cite{shalgar2022supernova, shalgar2022neutrino, nagakura2022grqknt, nagakura2022time}\footnote{Ref.~\cite{Stapleford:2019yqg} implemented flavor oscillations in spherically symmetric CCSN simulations, but taking the assumptions that flavor oscillations only occur much above the region where neutrinos decouple. Moreover, the flavor evolution history for all radially outgoing neutrinos were assumed to be independent of their propagation angles.}. 

In this work, we extend previous analyses and consider for the first time the collisional instability and its effects in spherically symmetric hydrostatic CCSN backgrounds.
Therefore, we solve the quantum kinetic transport equations using a multi-energy and multi-angle collective neutrino oscillation simulator, an extended version of \texttt{COSE$\nu$}~\cite{george2022cose}.
As spherically symmetric neutrino transport with realistic collisional rates do not lead to angular spectrum crossing and hence no fast instability~\cite{tamborra2017flavor}, it provides a clean background for us to probe the consequence of collisional flavor instability, which should exist in regions where the fast instabilities do not exist.
We consider several radial profiles of thermal quantities at selected post-bounce times, obtained from a CCSN simulation.
For each snapshot, we then simulate the evolution of neutrinos up to $\sim 1$~ms in our simulator including advection, collisions, and flavor oscillations based on state-of-the-art weak rates determined from the background CCSN profiles, however, without feedback on the medium.
We start with fiducial models by including minimal but essential ingredients that are able to trigger the collisional instability.
Additional ingredients are added case by case to explore their possible impact.

This paper is organized as follows.
In Sec.~\ref{sec:model}, we describe our models and the adopted parameters.
We present the linear stability analyses for collisional instabilities in Sec.~\ref{sec:linear}.
We discuss our simulation results of fiducial models in Sec.~\ref{sec:results} and those with additional parameters in Sec.~\ref{sec:effects}.
Further discussions and conclusions are given in Sec.~\ref{sec:discussion}.
We adopt natural units and $\hbar=c=k_B=1$ throughout the paper.

\section{Models}
\label{sec:model}
\subsection{CCSN simulation and neutrino collisional processes}
We use \texttt{AGILE-BOLTZTRAN} to simulate a CCSN, launched from a 18 $M_\odot$ progenitor star, based on general relativistic neutrino radiation-hydrodynamics in spherical symmetry \cite{mezzacappa1993numerical, mezzacappa1993type, mezzacappa1993stellar}.
A comoving baryon mass mesh is used for 208 radial grid points \cite{liebendorfer2001conservative, liebendorfer2004finite}, which features an adaptive mesh refinement method \cite{liebendorfer2002adaptive} and implements as coordinate the enclosed baryon mass instead of the actual mass mesh location \cite{Fischer:2010A&A517}.

For the current study, we use the nuclear equation of state of Ref.~\cite{hempel2010statistical}, which is based on the nuclear statistical equilibrium approach for the composition of nuclei based on several 1000 species, in combination with the density dependent DD2 relativistic mean-field model of Ref.~\cite{typel2010composition}.
For low temperatures, $T< 0.45$~MeV, the ideal silicon- and sulfur gas approximation is applied.
Electron, positron, photon, and Coulomb contributions are added following the equation of state of \cite{timmes2000accuracy}.
Note that in the current study muons are not included in the SN simulation.

The Boltzmann neutrino-transport module of \texttt{AGILE-BOLTZTRAN} employs the discrete ordinate method, and evolves four species of neutrinos, $\nu_e$, $\bar\nu_e$,  $\nu_x$ (for $\nu_\mu$ or $\nu_\tau$), and $\bar \nu_x$ (for $\bar\nu_\mu$ or $\bar\nu_\tau$) without coherent flavor oscillations.
Although several weak collisional processes associated with muons were employed recently \cite{guo2020charged, fischer2020muonization}, we only include the processes from (1a) to (4b) listed in Table~\ref{tab:nu_process}:
EA~\cite{Fischer:2020PhRvC101}, NNS~\cite{bruenn1985stellar,mezzacappa1993numerical}, neutrino-electron scattering (NES)~\cite{mezzacappa1993stellar}, and neutrino-pair reactions (NPR), including both leptonic process~\cite{bruenn1985stellar} and nucleon-nucleon bremsstrahlung~\cite{ThompsonBurrows:2001NuPhA688,Fischer:2016A&A593}.

\begin{table}[t]
    \centering
    \caption{\label{tab:nu_process} Set of weak processes considered in \texttt{BOLTZTRAN} (short as \texttt{B}) or \texttt{COSE$\nu$} (\texttt{C}), where $\nu$ and $\bar\nu$ are for all neutrino flavors and $N = n, p$.}
    \begin{tabular}{lccc}
    \hline\hline
    Label & Weak process & Abbreviation & Adoption \\\hline
    (1a) & $\nu_e + n \leftrightarrows p + e^-$ & EA & \texttt{B},\texttt{C} \\
    (1b) & $\bar\nu_e + p \leftrightarrows n + e^+$ & EA & \texttt{B},\texttt{C} \\
    (1c) & $\bar\nu_e + p + e^- \leftrightarrows n$ & EA & \texttt{B},\texttt{C}\vspace{0.05in}\\
    (2a) & $\nu + N \leftrightarrows \nu + N $ & NNS & \texttt{B},\texttt{C} \\
    (2b) & $\bar\nu + N \leftrightarrows \bar\nu + N $ & NNS & \texttt{B},\texttt{C}\vspace{0.05in}\\
    (3) & $\nu + e^{\pm} \leftrightarrows \nu + e^{\pm} $ & NES & \texttt{B}\vspace{0.05in}\\
    (4a) & $\nu + \bar\nu \leftrightarrows e^- + e^+ $ & NPR & \texttt{B} \\
    (4b) & $\nu + \bar\nu + N+N \leftrightarrows N+N $ & NPR & \texttt{B}\vspace{0.05in}\\
    (5a) & $\nu_\mu + n \leftrightarrows p + \mu^-$ & EA (muonic) & \texttt{C} \\
    (5b) & $\bar\nu_\mu + p \leftrightarrows n + \mu^+$ & EA (muonic) & \texttt{C} \\
    \hline\hline
    \end{tabular}
\end{table}

The roles of those collisional processes on the spectra of various neutrino species are different.
The energy spectra of $\nu_e$ and $\bar\nu_e$ near the neutrino sphere of last inelastic collision are predominantly determined by the EA processes
\footnote{Neutrinos with different energy decouple from matter at different radii; for details, c.f. Ref.~\cite{fischer2012neutrino}.}.
For $\nu_x$ and $\bar\nu_x$, they primarily interact with the medium through the neutral-current NNS.
The NNS are considered here in the elastic approximation, known as iso-energetic NNS.
In particular near the $\nu_x$ and $\bar\nu_x$ sphere of last elastic scattering, where the nucleon mass is much larger than neutrino energy, little energy exchange between neutrinos and medium happens.
Thus, the NNS only play important role in trapping $\nu_x$ and $\bar\nu_x$ and define their ``transport neutrino sphere'', which locates outside the ``energy sphere'', within which the NES and NPR processes thermalize $\nu_x$ and $\bar\nu_x$~\cite{bruenn1985stellar,raffelt2002mu,keil2003monte}.
For instance, when NES are included, the mean energy of $\nu_x$ and $\bar\nu_x$ outside their neutrino spheres can be as low as $\sim 60\%$ of the values obtained without including NES \cite{keil2003monte}.

The numerical implementation of EA and NNS processes are very different from NES and NPR, because the later two are inelastic in nature such that the scattering kernels connect both the energies and angles of incoming and outgoing neutrinos, which significantly increase the computational cost.
When more angular grids are used in, e.g., the simulations by \texttt{COSE$\nu$} (see next subsection), the computational complexity increases dramatically.
This prevents us from including NES and NPR unless approximated prescriptions for them are employed.
However, since these rates are subleading for $\nu_e$ and $\bar\nu_e$ whose rate difference is the key to trigger collisional instability, we expect that the results would remain qualitatively true when the NES and NPR rates are omitted.
We note that for $\nu_\mu$ and $\bar\nu_\mu$, if we only include the elastic NNS, there is no other process responsible for thermalizing them.
To partly compensating for this, we additionally include the muonic EA processes of (5a)-(5b) in Table~\ref{tab:nu_process} in extended \texttt{COSE$\nu$} simulations described in the next section, which can thermalize the spectrum of muonic neutrinos with energy $\gtrsim 65\, \mathrm{MeV}$ \cite{fischer2020muonization}.

\subsection{Neutrino transport including flavor oscillations and collisions}
\subsubsection{Quantum kinetic equation}
Although a full description of oscillation phenomena in CCSNe may require to consider three active neutrino flavors \cite{duan2008stepwise, dasgupta2008spectral, capozzi2020mu, Capozzi:2022dtr}, we take the approximation by considering only electron and muon flavors.
The general equations governing the spatial and temporal evolution of Wigner transformed density matrices $\varrho$ (for neutrinos) and $\bar\varrho$ (for anti-neutrinos), are given by
\begin{align}
    & ( \partial_t + v_r \partial_r + \frac{1-v_r^2}{r}\partial_{v_r}) \varrho(E, v_r, r, t) = \nonumber\\
    & -i[\mathbf H(E, v_r, r, t), \varrho(E, v_r, r, t)] +\mathbf C(E, v_r, r, t),
    \label{eq:eom_nu}
\end{align}
\begin{align}
    & ( \partial_t + v_r \partial_r +\frac{1-v_r^2}{r}\partial_{v_r}) \bar\varrho(E, v_r, r, t) = \nonumber\\
    & -i[\bar{\mathbf H}(E, v_r, r, t), \bar\varrho(E, v_r, r, t)] +\bar{\mathbf C}(E, v_r, r, t),
    \label{eq:eom_nubar}
\end{align}
with
\begin{align}
    \varrho(E, v_r, r, t) & = \begin{bmatrix} \varrho_{ee}(E, v_r, r, t) & \varrho_{e\mu}(E, v_r, r, t) \\ \varrho_{e\mu}^*(E, v_r, r, t) & \varrho_{\mu\mu}(E, v_r, r, t) \end{bmatrix}  , \nonumber\\
    \bar\varrho(E, v_r, r, t) & = \begin{bmatrix} \bar\varrho_{ee}(E, v_r, r, t) & \bar\varrho_{e\mu}(E, v_r, r, t) \\ \bar\varrho_{e\mu}^*(E, v_r, r, t) & \bar\varrho_{\mu\mu}(E, v_r, r, t) \end{bmatrix},
\end{align}
 in the flavor basis and normalized by the neutrino number density so that $n_{\nu_i}(r, t) = \int d E\, d v_r\, \varrho_{ii}(E, v_r, r, t)$ where $i=e,\,\mu$\footnote{The dependence on spacial and temporal indices will not be explicitly written unless emphasized in the following discussions.}.
Clearly, the diagonal elements $\varrho_{ii}$ relate to the neutrino phase-space distribution function $f_{\nu_i}$ by $\varrho_{ii}=f_{\nu_i} \times E^2/(2\pi^2)$ while the off diagonal elements characterize the degree of flavor mixing.
In Eqs.~\eqref{eq:eom_nu} and \eqref{eq:eom_nubar}, the advection term $[v_r \partial_r - (1/r)(1-v_r^2) \partial_{v_r}]$ takes a simplification from the full QKE \cite{blaschke2016neutrino,richers2019neutrino}, and the second part accounts for the aberration as the radial projection of velocity $v_r$ (the cosine of the angle between the traveling direction of neutrino and the radial direction) changes for neutrinos propagating non-radially.

On the right-hand side, $\mathbf H$ and $\mathbf C$ are the coherent propagation Hamiltonian and collisional term for neutrinos respectively.
Three different contributions to $\mathbf H$ include the vacuum mixing term
\begin{equation}
    \mathbf H_\mathrm{vac}(E) =
    \frac{\delta m^2}{4E}
    \begin{bmatrix}
    -\cos 2\theta_V & \sin 2 \theta_V \\ \sin 2 \theta_V & \cos 2 \theta_V
    \end{bmatrix},
\end{equation}
where $\delta m^2$ is the vacuum mass-squared difference and $\theta_V$ is the mixing angle, the diagonal matrix of the matter term
\begin{equation}
    \mathbf H_\mathrm{mat} = \mathrm{diag}[V_{\rm mat}, 0],
\end{equation}
with $V_{\rm mat}=\sqrt{2} G_F \rho Y_e/m_u$ corresponding to neutrino forward scattering on $e^{\pm}$, neutrons, and protons where $m_u$ is the atomic mass, and the neutrino self-induced term
\begin{align}
    \mathbf H_{\nu\nu}(v_r) = & \sqrt{2} G_F \int d E'\, d v_r'\, \times \nonumber\\
    & (1-v_r v_r') [\varrho(E',v_r')-\bar\varrho^*(E',v_r')],
\end{align}
corresponding to neutrino forward scattering on other neutrinos.

For the collisional term $\mathbf C$, we include the processes of EA and NNS in our calculations.
The contribution from EA is given by
\begin{align}
    \mathbf C_\mathrm{EA}(E) = &
    \frac{1}{2} \left\{ \mathrm{diag}[j_e(E), j_\mu(E)], \varrho_{\rm FO}(E)-\varrho(E) \right\} \nonumber\\
    & - \frac{1}{2} \left\{ \mathrm{diag}[\chi_e(E), \chi_\mu(E)], \varrho(E) \right\},
\end{align}
where $j_e(E)$ and $j_\mu(E)$ are the emissivities for the reactions for processes of (1a)-(1c) and (5a)-(5b) in Table~\ref{tab:nu_process}, $\chi_e(E)$ and $\chi_\mu(E)$ are the opacities for their inverse reactions, $\varrho_{\rm FO}(E)=\mathcal{I}\times E^2/(2\pi^2)$, with $\mathcal{I}$ being the identity matrix, is the fully-occupied differential number density for a specific $E$, and the curly bracket is the anti-commutator.
For the process of NNS, we have
\begin{align}
    \mathbf C_\mathrm{NNS}(E, v_r) = \int d v_r'\, & R_{\rm NNS}(E, v_r, v_r') \times \nonumber\\
    &  [\varrho(E, v_r')-\varrho(E, v_r)],
\end{align}
where $R_{\rm NNS}(E, v_r, v_r')$ is the scattering kernel transferring neutrino of energy $E$ with a radial velocity $v_r$ to the same energy but a different velocity $v_r'$.
The special feature of the scattering kernel in NNS, $R_{\rm NNS}(E, v_r, v_r') = \chi_{\rm NNS} (E)/2 + v_r v_r' \tilde\chi_{\rm NNS} (E)/2 $,
allows further simplification
\begin{align}
    \mathbf C_\mathrm{NNS}(E, v_r) = & -\chi_{\rm NNS} (E) \left[\varrho(E, v_r) - \int \frac{d v_r'}{2} \, \varrho(E, v_r')\right] \nonumber\\
    & + v_r \tilde\chi_{\rm NNS} (E) \int \frac{d v_r'}{2}\, v_r' \varrho(E, v_r'),
    \label{eq:CNNS_chiNNS}
\end{align}
where $\chi_{\rm NNS} (E)$ and $\tilde\chi_{\rm NNS} (E)$ are the opacity in NNS.
Notice that the Pauli blocking in iso-energetic NNS has no impact because $R_{\rm NNS}(E, v_r, v_r')=R_{\rm NNS}(E, v_r', v_r)$ as shown in Appendix~\ref{sec:A1}.

For anti-neutrinos, $\bar{\mathbf H}$ and $\bar{\mathbf C}$ on the right-hand side are defined similarly: $ \bar{\mathbf H}_\mathrm{vac}(E) = \mathbf H_\mathrm{vac}(E) $, $\bar{\mathbf H}_\mathrm{mat} = -\mathbf H_\mathrm{mat} $,  $ \bar{\mathbf H}_{\nu\nu}(v_r) = -\mathbf H^*_{\nu\nu}(v_r) $,
\begin{align}
    \bar{\mathbf C}_\mathrm{EA}(E) = &
    \frac{1}{2} \left\{ \mathrm{diag}[\bar j_e(E), \bar j_\mu(E)], \bar\varrho_{\rm FO}(E)-\bar\varrho(E) \right\} \nonumber\\
    & - \frac{1}{2} \left\{ \mathrm{diag}[\bar\chi_e(E), \bar\chi_\mu(E)], \bar\varrho(E) \right\},
\end{align}
and
\begin{align}
    \bar{\mathbf C}_\mathrm{NNS}(E, v_r) = \int d v_r'\, & \bar R_{NNS}(E, v_r, v_r') \times \nonumber\\
    & [\bar\varrho(E, v_r')-\bar\varrho(E, v_r)],
\end{align}
where $\bar j_e(E)$, $\bar j_\mu(E)$, $\bar\chi_e(E)$, $\bar\chi_\mu(E)$, and $\bar R_{\rm NNS}(E, v_r, v_r')$ are the emissivities, opacities, and scattering kernel for antineutrinos.

We use emissivities and opacities from \texttt{BOLTZTRAN} and calculate the kernel of NNS based on the method in Refs.~\cite{bruenn1985stellar, rampp2002radiation} for the simulations with \texttt{COSE$\nu$} (see Appendix~\ref{sec:A1} for details).

\subsubsection{Attenuation factors used in simulations}\label{sec:attenuation}
\begin{figure}[t]
\includegraphics[width=0.48\textwidth]{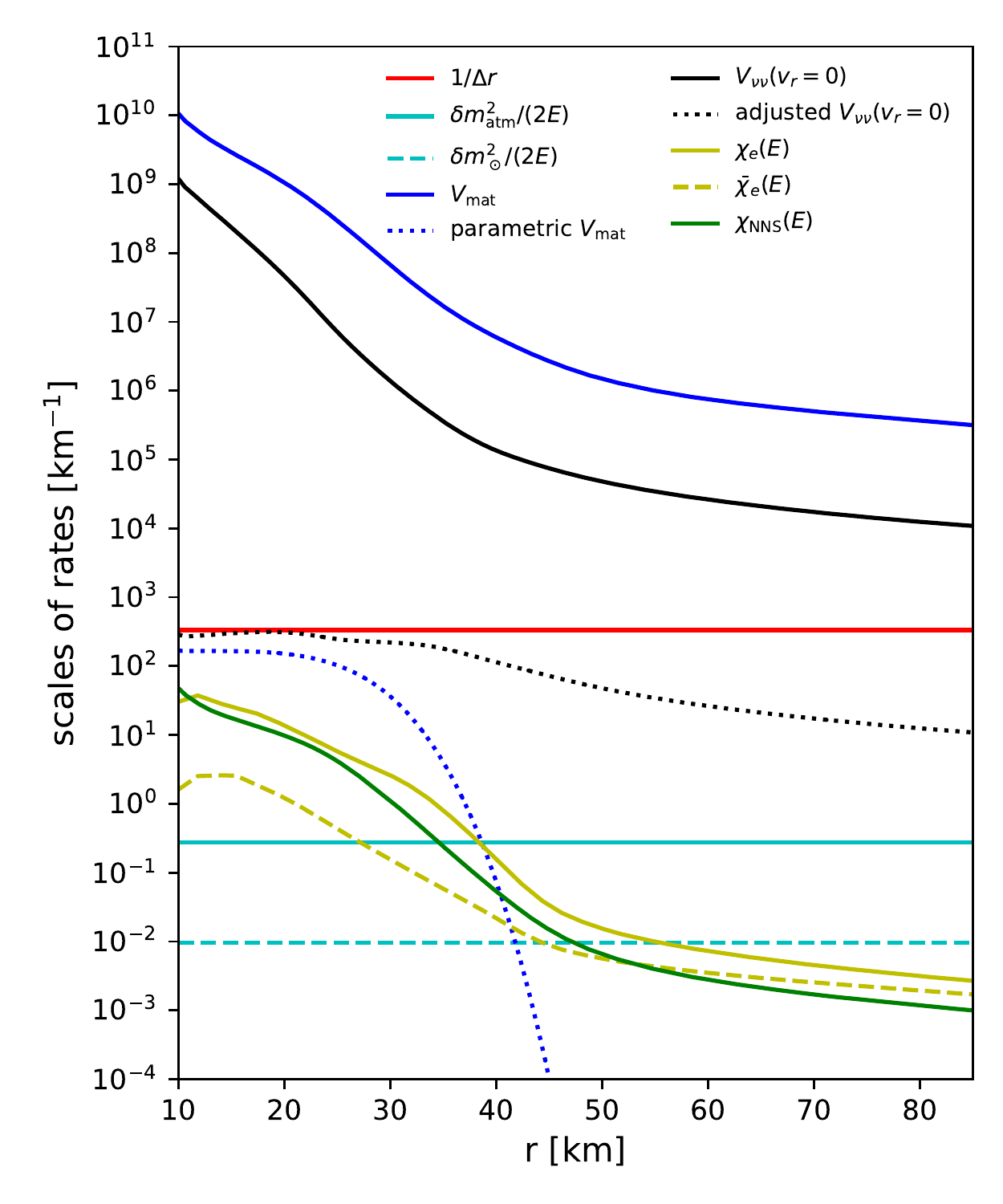}
\caption{\label{fig:rates_comparison} Comparison of the scales of all the included collision rates and different potentials in the Hamiltonian for neutrino energy $E=21$~MeV in the transport equation [Eqs.~\eqref{eq:eom_nu} and \eqref{eq:eom_nubar}] for the background snapshot at $t_\mathrm{pb} \approx 247$~ms (our Model II) from the CCSN simulation.
Also shown is $(\Delta r)^{-1}$ with the radial grid size $\Delta r=3$~m adopted in our simulations.
The black dotted curve shows the attenuated potential $V_{\nu\nu}$ (see Sec.~\ref{sec:attenuation} and Table~\ref{tab:parameters}) while the blue dotted curve shows the parametrized matter potential $V_{\rm mat}$ (see Sec.~\ref{sec:matter}).
We take $\delta m^2_{\rm atm}=2.3\times 10^{-3}~{\rm eV}^2$ and $\delta m^2_\odot=8\times 10^{-5}~{\rm eV}^2$ corresponding to the measured values in atmospheric and solar neutrino experiments respectively. 
}
\end{figure}

The rates involved in the flavor evolution equation span a wide range in magnitudes.
Due to the nature of weak interactions, the collisional rates are much lower than those of coherent forward scatterings by a factor of $\sim G_F E^2$ that can be $\sim 10^{-9}$--$10^{-5}$ for relevant neutrino energies.
For example, Fig.~\ref{fig:rates_comparison} compares the scales of all rates for neutrino energy $E=21\,\mathrm{MeV}$ computed at a post-bounce time $t_\mathrm{pb}\approx 247$~ms of the simulated CCSN.
The magnitude of neutrino self-induced term $\mathbf H_{\nu\nu}$ can be evaluated by taking the difference of its diagonal elements $V_{\nu\nu}= H_{\nu\nu,ee}-H_{\nu\nu,\mu\mu}$. 
It varies from $\sim 10^9~\mathrm{km}^{-1}$ at 10 km to $\sim 10^4~\mathrm{km}^{-1}$ at 85 km, while the opacities of $\nu_e$ or $\bar\nu_e$ are $\sim 10~\mathrm{km}^{-1}$ at 10 km and as low as $\sim 10^{-3}~\mathrm{km}^{-1}$ at 85 km.
This huge gap poses a computational challenge to properly handle all those rates.
One possible approach is to focus on a local simulation within $\mathcal O(10~\mathrm{m})$ or even smaller range \cite{martin2021fast}, but it loses the capability to account for global advection, and thus cannot model the long-term evolution of the system self-consistently.
Another constraint comes from the minimal radial interval $\Delta r$.
Simulating in a larger radial range $\sim \mathcal O(10~\mathrm{km})$ implies that the number of uniformly distributed radial grids $N_r=L/\Delta r$ in a radial range $L=75$~km must be $\gtrsim 10^8$ to properly resolve the length scale of oscillating flavor wave imposed by $\mathbf H_{\nu\nu}$ developed at $\sim 30$ km (see Sec.~\ref{sec:results}).
Taking such a large number of radial grids together with other required resolutions in angular and energy distributions is not feasible for our extended version of \texttt{COSE}$\nu$ at present.
Therefore, for the trilemma among self-consistency, advection, and exact rates, at least one component must be compromised.
For this work, we take the approach of introducing artificial attenuations to some of the rates as follows.

First, we scale down all elements in neutrino self-induced term $\mathbf H_{\nu\nu}$ by multiplying them with an attenuating factor \cite{nagakura2022grqknt},
\begin{equation}
  a_{\nu\nu}(r) = \frac{a_1}{1+e^{(a_2-r)/a_3}},
  \label{eq:attenuation_nu}
\end{equation}
where $a_1$, $a_2$, and $a_3$ are parameters, for which we take different values for different snapshot models (see Table~\ref{tab:parameters}). 
For instance, adopting $a_1=10^{-3}$, $a_2=35~{\rm km}$, and $a_3 = 3.0~{\rm km}$ for the snapshot shown in Fig.~\ref{fig:rates_comparison} results in attenuated $V_{\nu\nu}$ (black dotted curve) smaller than the value of $(\Delta r)^{-1}$, when taking $N_r=25000$ uniformly distributed radial grids.
This form attenuates $V_{\nu\nu}$ more at smaller radii as the neutrino number densities are significantly higher.
Notice that we also make sure our choice of $a_{\nu\nu}(r)$ always maintains the hierarchy between $V_{\nu\nu}$ and other collisional terms for most relevant neutrino energies. 
For example, the attenuated $V_{\nu\nu}$ shown in Fig.~\ref{fig:rates_comparison} remains larger than the collisional rates by at least a factor of 10--100 for neutrino energy of 21~MeV.

Second, for the matter potential, the high density around the neutrino sphere also leads to too large value of $V_{\rm mat}$ to be directly included in our simulation without modifications (see Fig.~\ref{fig:rates_comparison}).
Thus, for models that we include $\mathbf{H}_{\rm mat}$ and $\mathbf{\bar H}_{\rm mat}$, we instead adopt a parametrized function to explore the effect due to the presence of an inhomogeneous matter term (see Sec.~\ref{sec:matter}).

Third, the collisional rates for neutrinos with high energy $E\gtrsim 90$~MeV can exceed $(\Delta r)^{-1}$.
To avoid that, we also apply attenuation functions
\begin{equation}
	a_{\rm EA}(E, v_r) = \left[ 1+\frac{\chi_{\rm EA, max}^2(E)}{{a^2_4}} \right]^{-1/2},
\end{equation}
to all emissivities and opacities in EA and
\begin{equation}
	a_{\rm NNS}(E, v_r) = \left[ 1+\frac{\chi_{\rm NNS}^2(E)}{{a^2_5}} \right]^{-1/2},
\end{equation}
to the scattering kernels in NNS, where $\chi_{\rm EA,max}(E)$ is the maximal between $j_e(E) + \chi_e(E)$ and $j_\mu(E) + \chi_\mu(E)$, $a_4$ and $a_5$ are parametric saturation rates.
We adopt values of $a_4$ and $a_5$ close to $(\Delta r)^{-1}$ to make sure all rates are smaller than $(\Delta r)^{-1}$ for all energies. 
Note that for neutrinos with energy $\lesssim 90$~MeV, their rates are practically not attenuated since $\chi_{\rm EA, max}/a_4$
and $\chi_{\rm NNS}/a_5$ are much smaller than 1.
For those with $E\gtrsim 90$~MeV, they remain strongly trapped during our simulation duration. 

\subsubsection{Boundary and initial conditions}
We numerically solve the QKE under spherical symmetry within a radial range between an inner boundary $r_{i.b.}=10$~km and an outer boundary $r_{o.b.}=85$~km.
Right above the inner boundary, we set up a region with a length $\sim l_{\rm i.b.}$ and decaying width $w_{i.b.}$ wherein the EA rates for all energies are artificially increased by the amounts $\Delta j_i(E)$ and $\Delta \chi_i (E)$ as follows ($i=e,\mu$),
\begin{align}
	\Delta j_i(E) & = \frac{N_r/L}{1+e^{(r-r_{i.b.}-l_{i.b.})/w_{i.b.}}} \frac{\varrho_{{\rm eq},ii}(E)}{\varrho_{\rm FO}(E)+\varrho_{{\rm eq},ii}(E)}, \nonumber\\
	\Delta \chi_i(E) & = \frac{N_r/L}{1+e^{(r-r_{i.b.}-l_{i.b.})/w_{i.b.}}}\frac{\varrho_{\rm FO}(E)}{\varrho_{\rm FO}(E)+\varrho_{{\rm eq},ii}(E)},
\end{align}
before we apply the attenuation factor described in the above subsection, such that neutrinos in this zone reach equilibrium state $\varrho_{{\rm eq},ii}$ determined by the local temperature and the equilibrium neutrino chemical potential $\mu_\nu^{\rm (eq)}=\mu_e+\mu_p-\mu_n$ within $\sim 0.5$~ms.
The purpose of having this equilibrium zone is mainly to prevent the artificially fast leakage of 
$\bar\nu_e$ below $\sim 10$~MeV and $\nu_\mu$ below $\sim 20$~MeV due to the lack of their main production and opacity source of (inverse) bremsstrahlung process that are not included here~\cite{Hannestad:1998xx,keil2003monte}.
At the outer boundary, we employ the free-streaming boundary condition for forward propagating neutrinos with $v_r\geq 0$, and force neutrino density matrices to be zero for back propagating ones.
A different outer boundary condition was used in Ref.~\cite{nagakura2022grqknt} where neutrinos with equilibrium number density multiplied by a dilution factor are injected inward.
We also tested this boundary condition and found negligible impact to our results when the dilution factor is $< 10^{-4}$ given that we focus on the transport near the neutrino sphere rather than the possible halo effect \cite{cherry2012neutrino} at large radii.

Since we evolve the neutrino density matrices over static matter backgrounds provided by snapshots from \texttt{AGILE-BOLTZTRAN}, but use a slightly different set of neutrino reactions, we do not directly take the detailed neutrino distributions from \texttt{BOLTZTRAN} and evolve them with Eqs.~\eqref{eq:eom_nu} and \eqref{eq:eom_nubar}.
Instead, we adopt a two-step approach as follows.
First, we take the local thermal equilibrium state $\varrho_{\rm eq}$ and $\bar\varrho_{\rm eq}$ at each radius (with nonzero values in diagonal components only) as the initial condition, and evolve them with extended \texttt{COSE}$\nu$ by neglecting $\mathbf{H}$ and $\mathbf{\bar H}$.
Typically, $\varrho$ and $\bar\varrho$ reach a stationary state after a simulation time $t\sim 1$~ms.
The stationary state derived in this way automatically contains anisotropic angular distribution near and outside the neutrino sphere due to the advection and collisions.
Note that the EA and NNS rates are all evaluated based on a given background profile with the exception that now we assume the condition of $\mu_{\nu_\mu}=0$ in the beginning instead of $Y_\mu=0$ (i.e., $\mu_\mu=0$).
It is a good approximation to have $\mu_{\nu_\mu} \approx \mu_\mu \approx 0$ because initially $\nu_\mu$ and $\bar{\nu}_\mu$ are produced by pair processes and the same for $\mu^+$ and $\mu^-$.

After obtaining the stationary state distribution of $\varrho$ and $\bar\varrho$ without flavor oscillations, we take this as the initial condition for simulations fully taking into account all terms in Eqs.~\eqref{eq:eom_nu} and \eqref{eq:eom_nubar} for each snapshot.

When the vacuum term $\mathbf{H}_{\rm vac}$ is ignored, we apply an initial seed of neutrino flavor mixing using an Gaussian perturbation for the off-diagonal elements of $\varrho_\nu$ ($\bar\varrho_\nu$) to trigger the flavor instability 
\begin{align}
	&\frac{\varrho^{\rm pert}_{e\mu}(E, v_r)}{\varrho_{ee}(E, v_r)-\varrho_{\mu\mu}(E, v_r)}
= \frac{\bar\varrho^{\rm pert}_{e\mu}(E, v_r)}{\bar\varrho_{ee}(E, v_r)-\bar\varrho_{\mu\mu}(E, v_r)} \nonumber\\
    = & 10^{-3} \exp\left[ -\left(\frac{r~{\rm [km]}-47.5}{10}\right)^2 \right].
\end{align}
We also examine several cases that include $\mathbf{H}_{\rm vac}$ (see Sec.~\ref{sec:vacuumterm}).
For those cases, because $\mathbf{H}_{\rm vac}$ generates flavor mixing automatically, no artificial perturbations are given initially.

\subsubsection{List of evolved models}
We take four hydrodynamical snapshot profiles at different post-bounce times $t_{\rm pb}=144$~ms, 247~ms, 503~ms, and 1000~ms from the \texttt{BOLTZTRAN} SN simulation, labeled by Models I--IV in Table~\ref{tab:parameters}.
For all models, we evolve $\varrho$ and $\bar\varrho$ between $r_{\rm i.b.}=10$~km and $r_{\rm o.b.}=85$~km with the total radial range $L=75$~km. 
For each snapshot, we perform three simulations using different radial resolutions.
Our fiducial models listed in Table~\ref{tab:parameters} are high resolution ones with $N_r=25000$. 
Comparison to results obtained with smaller $N_r=10000$ and 2500 are given in Appendix~\ref{sec:A2}.
For angular grids, we take $v_r$ uniformly between $-1$ and $1$ with $N_{v_r}=50$.
We have also examined cases with $N_{v_r}=200$ for the lowest radial resolution runs and found that the results are nearly identical (see Appendix~\ref{sec:A2}).
We take $N_E=20$ energy grids between 2~MeV to 160~MeV spaced nearly uniformly in logarithmic scale.

Since our main goal is to examine the flavor conversions triggered by the collisional instability without the interference of other types of flavor instability, we first only include in our fiducial models the neutrino self-induced term, the full form of EA, and the diagonal elements of $\mathbf{C}_{\rm NNS}$ for the NNS collisions, which are the minimal ingredients to induce the collisional instability.
For Model II, we perform additional simulations including $\mathbf{H}_{\rm vac}$ and $\mathbf{H}_{\rm m}$ separately and study their effects. 
Moreover, we introduce two auxiliary parameters $b_{\rm EA}$ and $b_{\rm NNS}$, which are multiplied to all off-diagonal elements of $\mathbf C_{\rm EA}$ and $\mathbf C_{\rm NNS}$. 
Different values of $b_{\rm EA}=1,2,4$ and $b_{\rm NNS}=0,1$ are taken to test possible impact of the size of the off-diagonal elements in $\mathbf C_{\rm EA}$ and $\mathbf C_{\rm NNS}$ on the results.
Table~\ref{tab:parameters} summarizes all models presented in this paper. 
Each simulation takes $\sim\mathcal O(30000)$ CPU hours for a duration of simulation time $t\sim 1$~ms.

\begingroup
\begin{table*}[t]
\begin{ruledtabular}
    \centering
    \caption{\label{tab:parameters} Parameters used in each model. For all models, the radial range is from 10~km to 85~km, $l_{\rm i.b.}=6$~km, $w_{i.b.}=0.2$~km, $N_r=25000$, $N_{v_r}=50$, $N_E=20$, $a_4 = 1/\Delta r$~[km$^{-1}$] , $a_5 = 1/(2\Delta r)$~[km$^{-1}$] and $\theta_V=10^{-6}$.}
    \begin{tabular}{cc ccc cccc}
        model & $t_\mathrm{pb}$ [ms] & $a_1$ & $a_2~\mathrm{[km]}$ & $a_3~{\rm [km]}$ & $b_\mathrm{EA}$ & $b_\mathrm{NNS}$ & $\delta m^2$ [eV$^{2}$] & $V_\mathrm{mat}(r~\mathrm{[km]})$~[km$^{-1}$] \\\hline
        I & 144 & $10^{-3}$ & 45 & 4.5 & 1 & 0 & 0 & 0 \\
        II & 247 & $10^{-3}$ & 35 & 3 & 1 & 0 & 0 & 0 \\
        III & 503 & $10^{-3}$ & 25 & 1.5 & 1 & 0 & 0 & 0 \\
        IV & 1000 & $10^{-3}$ & 21 & 1.1 & 1 & 0 & 0 & 0 \\\hline
        IIn & 247 & $10^{-3}$ & 35 & 3 & 1 & 1 & 0 & 0 \\
        IIe1 & 247 & $10^{-3}$ & 35 & 3 & 2 & 0 & 0 & 0 \\
        IIe2 & 247 & $10^{-3}$ & 35 & 3 & 4 & 0 & 0 & 0 \\
        IIv1 & 247 & $10^{-3}$ & 35 & 3 & 1 & 0 & $8\times 10^{-5}$ & 0 \\
        IIv2 & 247 & $10^{-3}$ & 35 & 3 & 1 & 0 & $2.3\times 10^{-3}$ & 0 \\
        IIm & 247 & $10^{-3}$ & 35 & 3 & 1 & 0 & 0 & $(500/3) \cdot\exp\left[ -(r-10)^4/18^4 \right]$ \\
    \end{tabular}
\end{ruledtabular}
\end{table*}
\endgroup

\section{Linear stability analysis}
\label{sec:linear}
We perform the linear stability analysis in the regime where the off-diagonal elements of density matrices are small and can be treated as perturbations compared to the number densities, i.e., $|\varrho_{e\mu}|/|\varrho_{ee}-\varrho_{\mu\mu}|\ll 1$ and $|\bar\varrho_{e\mu}|/|\varrho_{ee}-\varrho_{\mu\mu}|\ll 1$ \cite{banerjee2011linearized,izaguirre2017fast}.
Neglecting the vacuum and matter Hamiltonian contribution as well as the aberration term, we further assume that a collective mode of the perturbation $\varrho_{e\mu} = Q(\Omega,K_r,E,r,v_r)e^{-i[\Omega t-K_r (r'-r)]}$ and $\bar\varrho_{e\mu}^* = \bar Q(\Omega,K_r,E,r,v_r)e^{-i[\Omega t-K_r (r'-r)]}$ can develop locally near $r$.
With these assumptions, the off-diagonal parts of Eqs.~\eqref{eq:eom_nu} and \eqref{eq:eom_nubar} become
\begin{align}\label{eq:LEQ1}
& \left[ \Omega - K_r v_r - \Phi(v_r) + i C_{\rm e\mu,EA}(E) \right]  Q(\Omega,K_r,E,v_r) = \nonumber\\
& -\sqrt{2}G_F[\varrho_{ee}(E,v_r)-\varrho_{\mu\mu}(E,v_r)] \int dE'\, dv_r'\,\times \nonumber\\
&  (1- v_r v_r') [Q(\Omega,K_r,E',v_r')-\bar Q(\Omega,K_r,E',v_r')],
\end{align}
and
\begin{align}\label{eq:LEQ2}
& \left[\Omega - K_r v_r - \Phi(v_r) + i\bar C_{\rm e\mu,EA}(E) \right]  \bar Q(\Omega,K_r,E,v_r) = \nonumber\\
& -\sqrt{2}G_F[\bar\varrho_{ee}(E,v_r)-\bar\varrho_{\mu\mu}(E,v_r)] \int dE'\, dv_r'\,\times \nonumber\\
& (1-v_r v_r') [Q(\Omega,K_r,E',v_r')-\bar Q(\Omega,K_r,E',v_r')],
\end{align}
where $\Phi(v_r)=\sqrt{2}G_F\int dE'\, dv_r'\, (1-v_r v_r') [\varrho_{ee}(E',v_r')-\varrho_{\mu\mu}(E',v_r')-\bar\varrho_{ee}(E',v_r')+\bar\varrho_{\mu\mu}(E',v_r')]$, $C_{\rm e\mu,EA}(E)=[j_e(E)+\chi_e(E)+j_\mu(E)+\chi_\mu(E)]/2$, and $\bar C_{\rm e\mu,EA}(E)=[\bar j_e(E)+\bar \chi_e(E)+\bar j_\mu(E)+\bar \chi_\mu(E)]/2$.
The eigenvalues of $\Omega$ can be numerically derived for a given wave number $K_r$ using the same discretization scheme for both $E$ and $v_r$ as in the simulations. 
Solutions derived in this way will be shown in the next section to help understand the simulation outcome.

Another useful simplification that one can take is to consider the limit where the environment is homogeneous and all neutrinos are monochromatic.
Taking these assumptions, the linearized equations above can be further simplified into 
\begin{align}
( \Omega + i C_{\rm e\mu,EA} )  Q(\Omega) = &
- \sqrt{2}G_F [\Delta n_{\bar\nu} Q(\Omega) - \Delta n_{\nu} \bar Q(\Omega)], \nonumber\\
(\Omega  + i\bar C_{\rm e\mu,EA} )  \bar Q(\Omega) = & -\sqrt{2}G_F [\Delta n_{\bar\nu} Q(\Omega) - \Delta n_{\nu}\bar Q(\Omega)].
\end{align}
The corresponding eigenvalue of $\Omega$ therefore satisfies a quadratic secular equation
\begin{align}
 & \Omega^2 +[i(C_{\rm e\mu,EA} + \bar C_{\rm e\mu,EA}) + \sqrt{2}G_F(\Delta n_{\bar\nu}-\Delta n_{\nu})] \Omega \nonumber\\
= & i \sqrt{2}G_F(C_{\rm e\mu,EA} \Delta n_{\nu} - \bar C_{\rm e\mu,EA} \Delta n_{\bar\nu}) + C_{\rm e\mu,EA} \bar C_{\rm e\mu,EA},
\end{align}
where $\Delta n_{\nu}=n_{\nu_e}-n_{\nu_\mu}$ and $\Delta n_{\bar\nu}=n_{\bar\nu_e}-n_{\bar\nu_\mu}$.
The imaginary part of analytical solutions can be obtained as
\begin{align}
	{\rm Im}&(\Omega) \approx - \frac{|C_{\rm e\mu,EA} + \bar C_{\rm e\mu,EA}|}{2}  \pm \frac{|C_{\rm e\mu,EA} - \bar C_{\rm e\mu,EA}|}{2}  \times \nonumber \\
   & \frac{ |\Delta n_{\nu}+\Delta n_{\bar\nu}|}{|\Delta n_{\nu}-\Delta n_{\bar\nu}|}
    \left [ 1-\frac{1}{4 G_F^2} \left( \frac{C_{\rm e\mu,EA} - \bar C_{\rm e\mu,EA}}{\Delta n_{\nu}-\Delta n_{\bar\nu}} \right)^2 \right]
    \label{eq:LSA_Im}
\end{align}
to the order of $\mathcal O[C^3_{\rm e\mu,EA} /(G_F \Delta n_{\nu})^2]$.
When $\sqrt{2} G_F (\Delta n_\nu - \Delta n_{\bar\nu})$, i.e. $V_{\nu\nu}$, becomes much larger than the collisional rates, the high-order term is negligible and approaches to an asymptotic value.
Although this simplified expression cannot replace Eqs.~\eqref{eq:LEQ1} and \eqref{eq:LEQ2} to precisely predict the growth rates, it can help us quantitatively understand the general feature of the neutrino system.
Whether there exists a runaway solution roughly depends on the relation between the asymmetry factor of EA rates $\alpha_C\equiv |C_{\rm e\mu,EA} - \bar C_{\rm e\mu,EA}|/|C_{\rm e\mu,EA} + \bar C_{\rm e\mu,EA}|$  and that of neutrino number densities $\alpha_n\equiv |\Delta n_{\nu}-\Delta n_{\bar\nu}|/|\Delta n_{\nu}+\Delta n_{\bar\nu}|$~\cite{johns2021collisional}. 
Clearly, when ignoring higher order terms in Eq.~\eqref{eq:LSA_Im}, larger asymmetry in the EA rates and smaller asymmetry in neutrino number densities can result in a positive imaginary value of $\Omega$, which leads to an instability.
The criteria of having the instability is  usually satisfied near the neutrino sphere.
The reason is twofold.
First, around the neutrino sphere the $Y_e$ is usually low ($Y_e\lesssim 0.15$) such that the $\nu_e$ EA rates can be much larger than that of $\bar\nu_e$ by a factor of $\approx 5$--10.
Second, the $\nu_\mu$ and $\bar\nu_\mu$ number densities are usually higher than that of $\bar\nu_e$ well inside the neutrino sphere but become smaller around the sphere, due to the diffusion of $\nu_\mu$. 
At radii where $n_{\nu_\mu}\simeq n_{\bar\nu_\mu}\lesssim n_{\bar\nu_e}$, the asymmetry factor in neutrino number densities becomes the largest.
Thus, a combination of these two facts allows the instability generally exist around the neutrino sphere.

We caution that the factors discussed above are not necessarily the only ones determining how the collisional instability to induce flavor conversions.
This is because in the stability analysis presented above, we ignored the aberration term and assume the neutrino number density is locally homogeneous.
To understand the role of these dynamical properties, it requires a complete simulation as we will discuss in the next two sections.
However, the linear stability analysis still provides some insights to the underlying mechanism that drives the flavor instability as will be discussed further below.

\section{Results of collisional instability}
\label{sec:results}
For all four fiducial models (I--IV) that represent different evolution stages of the CCSN model, we observe the occurrence of collisional instability. 
The features in Model II, III, and IV are qualitatively similar while Model I shows a unique behavior.
Below, we begin by discussing results obtained in Model II.
We then address the similarities and differences of the models.

For this section, we define energy-integrated density matrices $\langle\varrho\rangle_E(v_r) = \int d E\, \varrho(E, v_r) $, angle-integrated density matrices $\langle\varrho\rangle_A(E) = \int d v_r\, \varrho(E, v_r)$, and neutrino mean energy
\begin{equation}
	\langle E_{\nu_i} \rangle = \frac{\int d E\,d v_r\, E \varrho_{ii}(E,v_r)}{\int d E\,d v_r\, \varrho_{ii}(E,v_r)},
\end{equation}
where $i=e,\,\mu$ for analysis.

\subsection{Model II}
\begin{figure*}[t]
\includegraphics[width=0.99\textwidth]{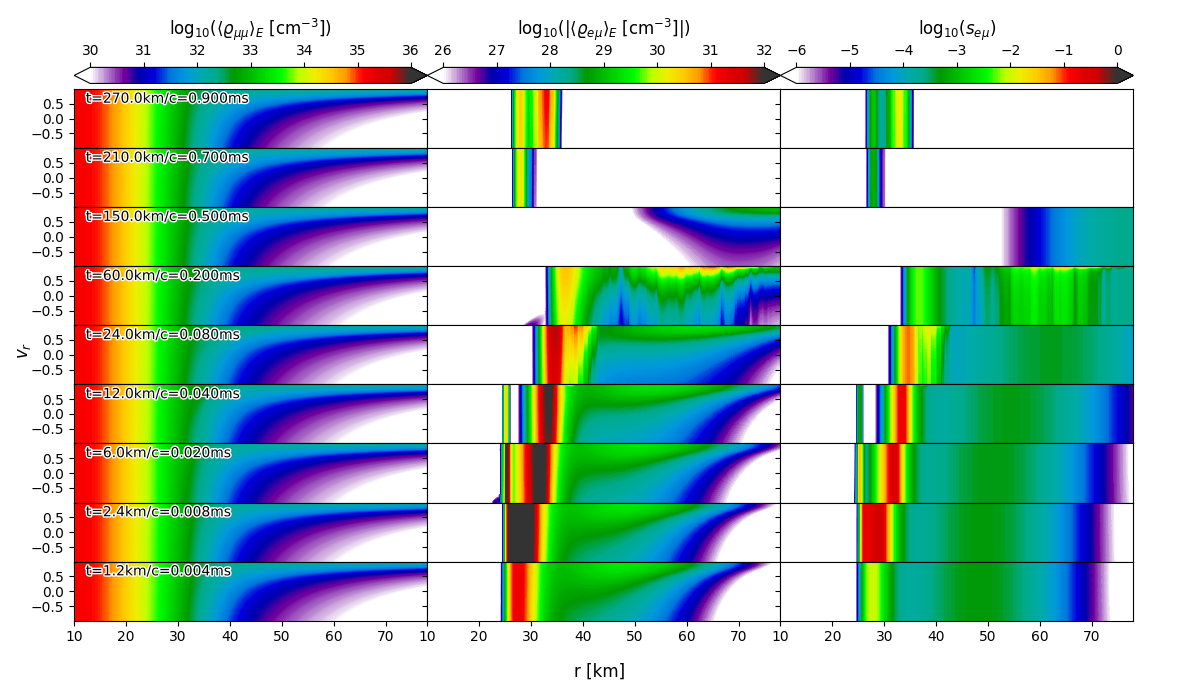}
\caption{\label{fig:2d_II} Evolution of energy-integrated number density $\langle \varrho_{\mu\mu} \rangle_E$ (left panel), flavor mixing $|\langle \varrho_{e\mu} \rangle_E|$ (middle panel), and the dimensionless ratio $s_{e\mu}\equiv |\langle \varrho_{e\mu} \rangle_E|/|\langle \varrho_{ee} \rangle_E-\langle \varrho_{\mu\mu} \rangle_E|$  (right panel) for Model II.}
\end{figure*}

We first show in Fig.~\ref{fig:2d_II} the energy-integrated diagonal and off-diagonal elements of neutrino density matrices $\langle \varrho_{\mu\mu} \rangle_E$ (left panels) and $|\langle \varrho_{e\mu} \rangle_E|$ (middle panels), as well as the dimensionless ratio $s_{e\mu} \equiv |\langle \varrho_{e\mu} \rangle_E|/|\langle \varrho_{ee} \rangle_E-\langle \varrho_{\mu\mu} \rangle_E|$ (right panels), as functions of $r$ and $v_r$ taken at different times from the \texttt{COSE$\nu$} simulation for Model II. 
Both the left and middle panels show the anisotropic shape of $\langle\varrho_{\mu\mu}\rangle$ and $|\langle\varrho_{e\mu}\rangle|$ at larger radii due to the nature of neutrino free streaming. 
However, the ratio $s_{e\mu}$ in right panels reveals that the development and evolution of the flavor off-diagonal elements in fact depend on $v_r$ rather weakly.
Clearly, the flavor instability existing around $r=25$-30~km in the beginning leads to the rapid growth of $|\langle \varrho_{e\mu} \rangle_E|$ to the nonlinear regime within $\sim 0.008$~ms, indicated by $s_{e\mu} \gtrsim 0.1$.
Afterwards, $s_{e\mu}$ stops growing in magnitude and splits into two branches that move in two opposite directions.
The inward moving one gets damped by the interaction with denser medium and eventually disappears at $\sim 0.04$~ms.
The outward going branch propagates with a group velocity of $\sim 0.4$ and further bifurcates into another two sub-branches at $\sim 36$~km around $t\sim 0.08$~ms.
The inner sub-branch moves at a similar group velocity of $\sim 0.4$ and disappears at $t\sim 0.5$~ms.
On the other hand, the outer one propagates with a group velocity of $\sim 0.9$, quick enough to reach the free-streaming region where flavor mixing gets transported away from our simulation domain for positive $v_r$ modes.
Afterwards, the neutrino fields remain stable (against oscillations) for a duration of $\sim 0.2$~ms during 0.5~ms$\lesssim t \lesssim 0.7$~ms.
At $t \approx 0.7$~ms, the collisional instability appears once again around $r\sim 28$~km after the neutrino field self-regulates its distributions near the decoupling region.
Unlike the first instability discussed above, the flavor mixing due to the second collisional instability does not get transported away this time and eventually freezes into a stationary state until the end of our simulation at $t\sim 1$~ms.

\begin{figure}[t]
\includegraphics[width=0.45\textwidth]{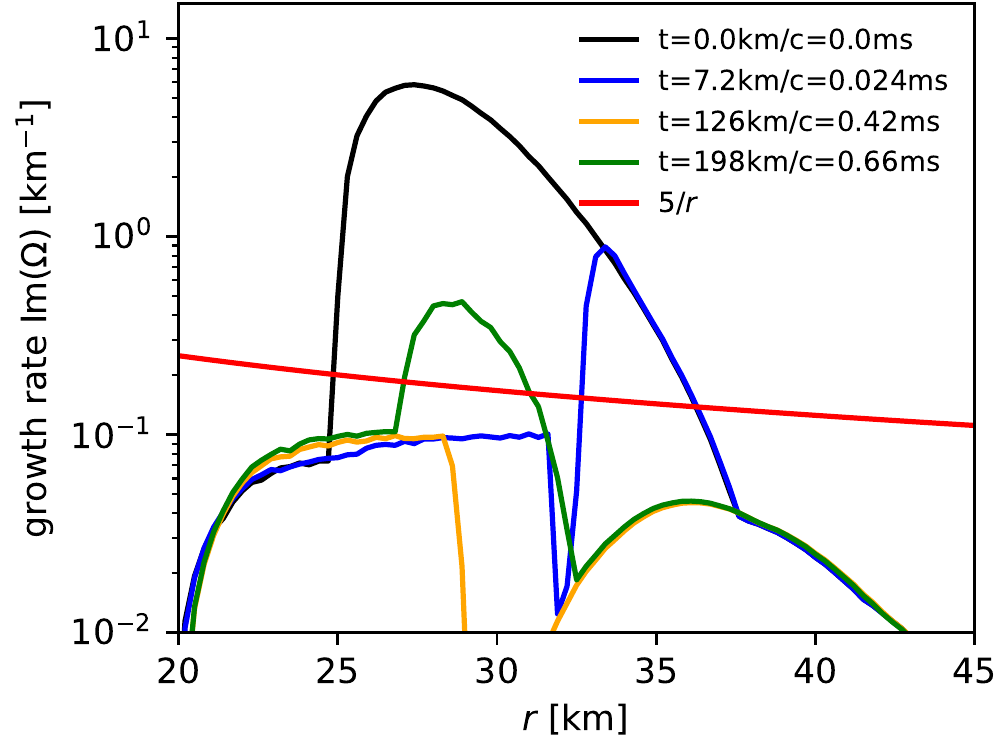}
\caption{\label{fig:LSA} Growth rates Im$(\Omega)$ from linear stability analysis as functions of radius $r$ for four simulation times in Model II. We sample 31 values of $K_r$ from -2~km$^{-1}$ to 2~km$^{-1}$ and show the maximal values for each radius. The red curve indicates $5/r$ as an empirical criteria to determine the growth of instability against advection.
}
\end{figure}

\begin{figure}[t]
\includegraphics[width=0.45\textwidth]{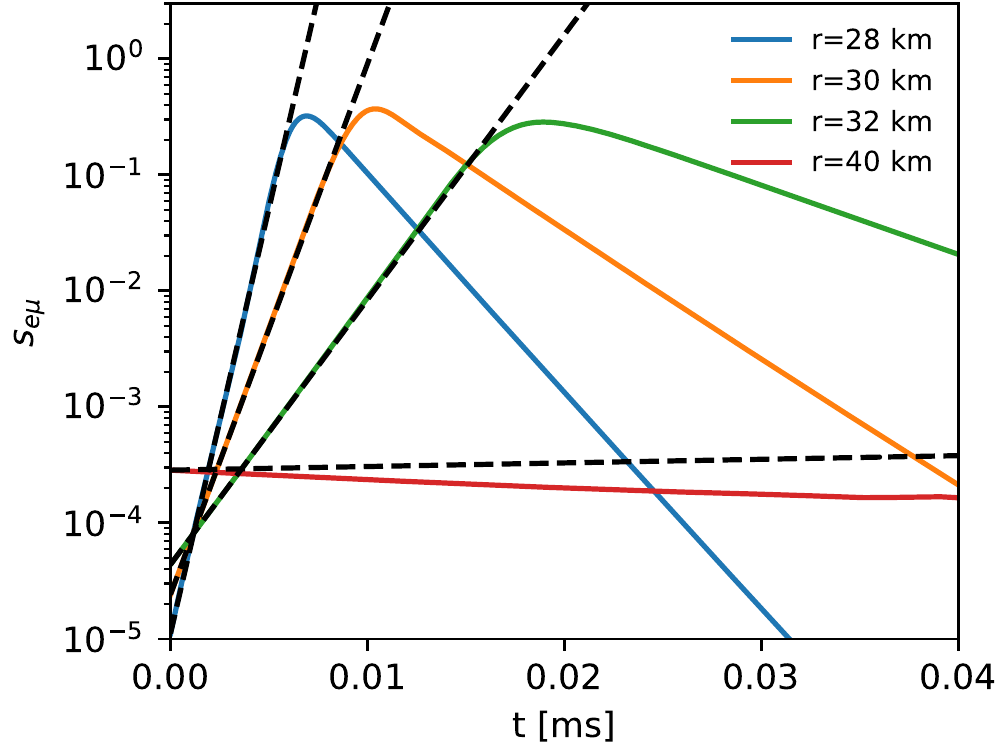}
\caption{\label{fig:time_evol} Time evolution of the dimensionless ratio $s_{e\mu}$ of radial velocity $v_r=1$ for four different radii in Model II. Each of them is compared with a black dashed line determined by $s_{e\mu}(t=0) \exp[{\rm Im}(\Omega) t]$ with the growth rate Im$(\Omega)$ in Fig.~\ref{fig:LSA} at $t=0$~ms.
}
\end{figure}

Figure~\ref{fig:LSA} shows the maximum growth rates of flavor instabilities among all $K_r$ modes, Im$(\Omega)$, as a function of radius for Model II at different times.
These Im$(\Omega)$ are derived by numerically solving the linearized Eqs.~\eqref{eq:LEQ1} and \eqref{eq:LEQ2}.
At $t=0$~ms, Im$(\Omega)$ peaks at $r\simeq 27$~km. 
The growth of this instability thus dominates the evolution of the system initially, consistent with results shown in Fig.~\ref{fig:2d_II}. 
When $t=0.024$~ms after flavor transformation occurs, Im$(\Omega)$ in 25~km$\lesssim r \lesssim32$~km becomes smaller than $\sim 0.1$~km$^{-1}$, while its value maintain roughly the same for $r\gtrsim 32$~km.
For $t=0.42$~ms when flavor mixings around $r\sim 30$~km gets suppressed, Im$(\Omega)<0.1$~km$^{-1}$ for all radii.
At an even later time $t=0.66$~ms when flavor conversion reappears (see Fig.~\ref{fig:2d_II}), larger Im$(\Omega)$ are found in 27~km$\lesssim r \lesssim31$~km again.

We compare the maximum growth rates obtained from the stability analysis at 0~ms with the numerical evolution for Model II at four different radii.
Figure~\ref{fig:time_evol} shows that the time evolution of $s_{e\mu}$ of radial velocity $v_r=1$ in the linear regime perfectly agree with the prediction determined by $s_{e\mu}(t=0) \exp[{\rm Im}(\Omega) t]$ at 28, 30, and 32~km respectively, which is expected since the growth of collisional instability dominates over the disturbance from advection.

Although positive Im$(\Omega)$ are found for nearly all radii larger than 20~km at all times, not all of them lead to the growth of $|\langle \varrho_{e\mu}\rangle|$ in the simulation. 
This is because the stability analyses can only tell how a perturbation evolves around where the local condition can be maintained.
However, in realistic simulations where advection occurs in the presence of inhomogenous neutrino number density, the instability growth rate needs to compete with advection for a perturbation to grow before it being transported away. 
To illustrate this, we show in Fig.~\ref{fig:2d_II} a characteristic value of advection rate as $5/r$ for Model II by the red solid line.
We take this function in an empirical way motivated by the advection term in Eqs.~\eqref{eq:eom_nu} and \eqref{eq:eom_nubar} being generally proportional to $1/r$.
Comparing Figs.~\ref{fig:2d_II} and \ref{fig:LSA}, it seems to suggest that when Im$(\Omega)$ is roughly less than $5/r$, the growth rate of the instability is too small against the advection, such that no significant flavor conversion can develop.
For example, although the stability analysis yields a positive Im$(\Omega)$ at $r=40$~km and $t=0$~ms, $s_{e\mu}$ decreases in the simulation as shown in Fig.~\ref{fig:time_evol}.
In addition, Im$(\Omega)$ are positive from 0.024~ms to 0.42~ms at $r = 27$~km in Fig.~\ref{fig:LSA}, while $s_{e\mu}$ decreases in Fig.~\ref{fig:2d_II}.

\begin{figure}
\subfigure{\includegraphics[width=0.45\textwidth]{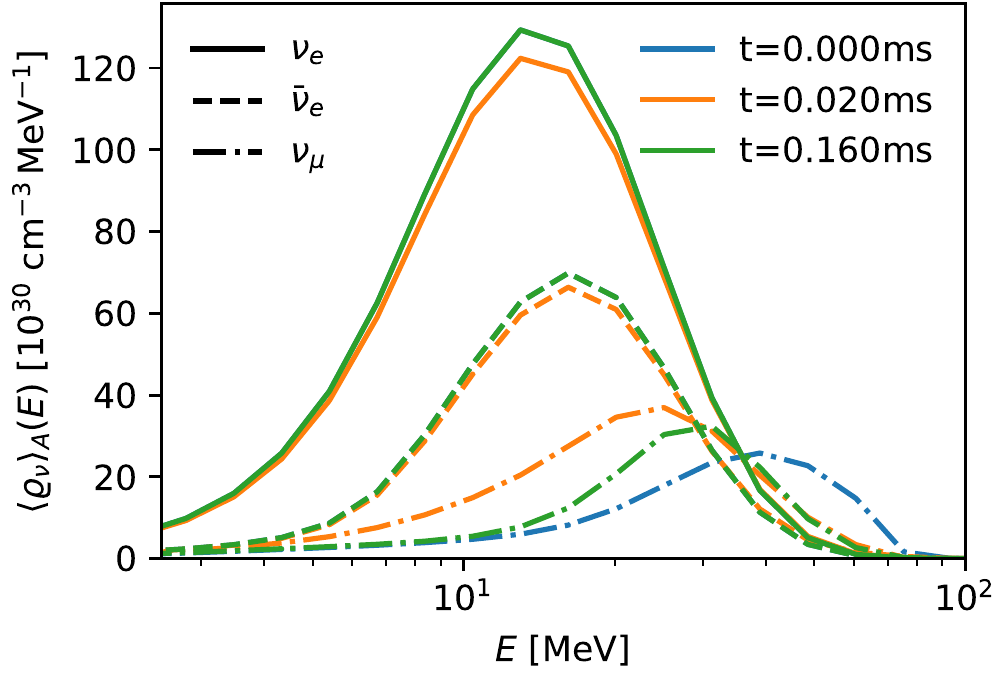}
\llap{\parbox[b]{4.1in}{\small (a)\\\rule{0ex}{1.9in}}} }
\subfigure{\includegraphics[width=0.45\textwidth]{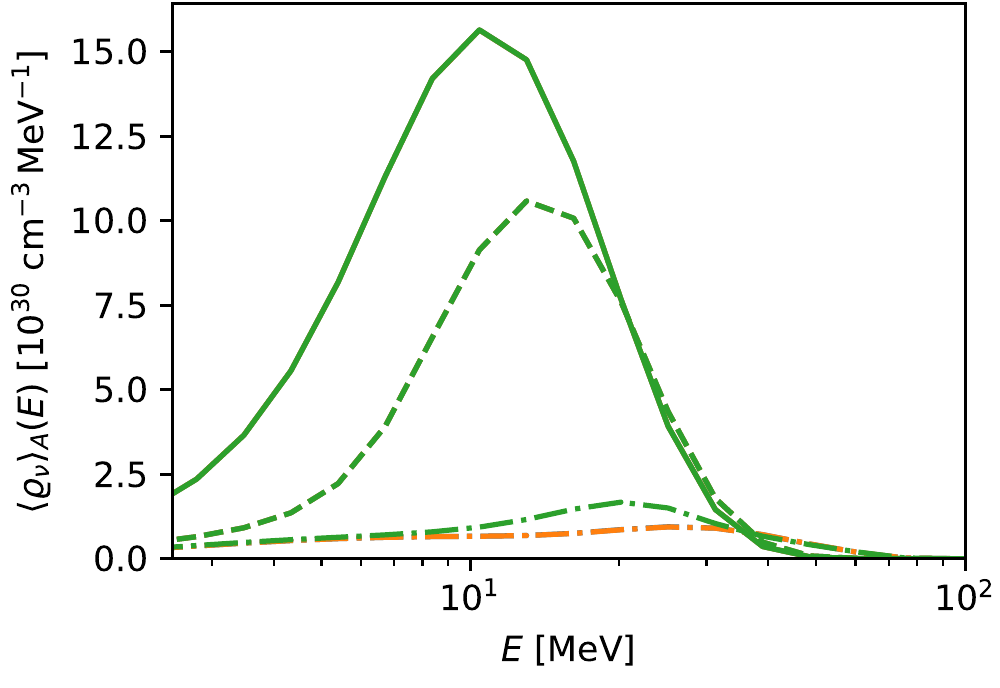}
\llap{\parbox[b]{4.9in}{\small (b)\\\rule{0ex}{1.9in}}} }
\caption{\label{fig:en_II} Angle-integrated energy spectrum for $\nu_e$(solid), $\bar\nu_e$ (dashed), and $\nu_\mu$ (dash-dotted) at two radii [32~km in panel (a) and 55~km in panel (b)]. Blue, orange, and green colors correspond to three different simulation times 0.0~ms, 0.02~ms, and 0.16~ms respectively. Note that in panel (b), the blue curve of overlaps with the orange curve for $\nu_\mu$, while the blue and orange curves overlap with the green curve for $\nu_e$ and $\bar\nu_e$.
}
\end{figure}

Next, we examine the impact of flavor conversion due to collisional instability on the property of neutrinos of all flavors.
Figure~\ref{fig:en_II} shows the angular-integrated neutrino energy spectra $\langle\varrho_\nu\rangle_A$ for two radii $r=32$ and $55$~km at three different simulation times $t=0.0$, 0.02, and $0.16$~ms.
For $r=32$~km close to where the initial flavor instability occurs, the maximum amount of flavor conversion happens when $t\sim 0.02$~ms. 
At this time, the $\nu_\mu$ spectrum is significantly shifted towards low energy compared to the one at $t=0$, because high-energy $\nu_\mu$ are converted to $\nu_e$ while low-energy $\nu_\mu$ receive contribution from flavor conversion of $\nu_e$ in return.
However, the spectra of both $\nu_e$ and $\bar\nu_e$ are only affected marginally (reduced by $\sim 5\%$--$8\%$ near 15~MeV).
This is mainly because the relatively large EA processes [(1a)-(1c) in Table~\ref{tab:nu_process}] that act on $\nu_e$ and $\bar\nu_e$ quickly bring them to a state close to thermal equilibrium.
At later times after the flavor mixing gets transported to outer radii, e.g., $t=0.16$~ms shown in Fig.~\ref{fig:en_II}, the EA processes restore the $\nu_e$ and $\bar\nu_e$ spectra fully back to the equilibrium state, while the advection of $\nu_\mu$ from denser region gradually shifts the peak of its spectrum to higher energy again.
For spectra at a larger radius of $r=55$~km, the $\nu_\mu$ spectrum remains identical as the initial one in the beginning (e.g., at $t=0.02$~ms) since there is no flavor instability at this radius. 
Differences only show up at later times (e.g., at $t=0.16$~ms) when the $\nu_\mu$ affected by collisional instability propagate here from inner radii, leading to a clearly visible shift of $\nu_\mu$ spectra to low energy. 
For $\nu_e$ and $\bar\nu_e$, EA collisions wipe out all the effects due to flavor conversions such that  their spectra remain unaffected by collisional instability at all.

Our results here suggest that there exists a fundamental difference in models assuming spatial homogeneity (e.g., Ref.~\cite{johns2021collisional}) and more realistic ones including advection.
For a model where spatial homogeneity is assumed and advection is neglected, flavor conversion due to collisional instability can lead to $\nu_\mu$ spectrum identical to that of $\nu_e$ or $\bar\nu_e$ given enough time.
This is because in homogeneous models, $\nu_\mu$ (decoupled from medium collisionally) continuously acquires supply from electron neutrinos due to flavor conversion while electron flavors can be repopulated by EA processes, so that their spectra becomes identical given long enough time.
However, similar to what argued above regarding the growth of instability, in a realistic environment, the advection introduces another timescale so that the transport of $\nu_\mu$ away from where the instability occurs can prevent their spectra become identical to that of $\nu_e$.

\begin{figure}[t]
\subfigure{\includegraphics[width=0.45\textwidth]{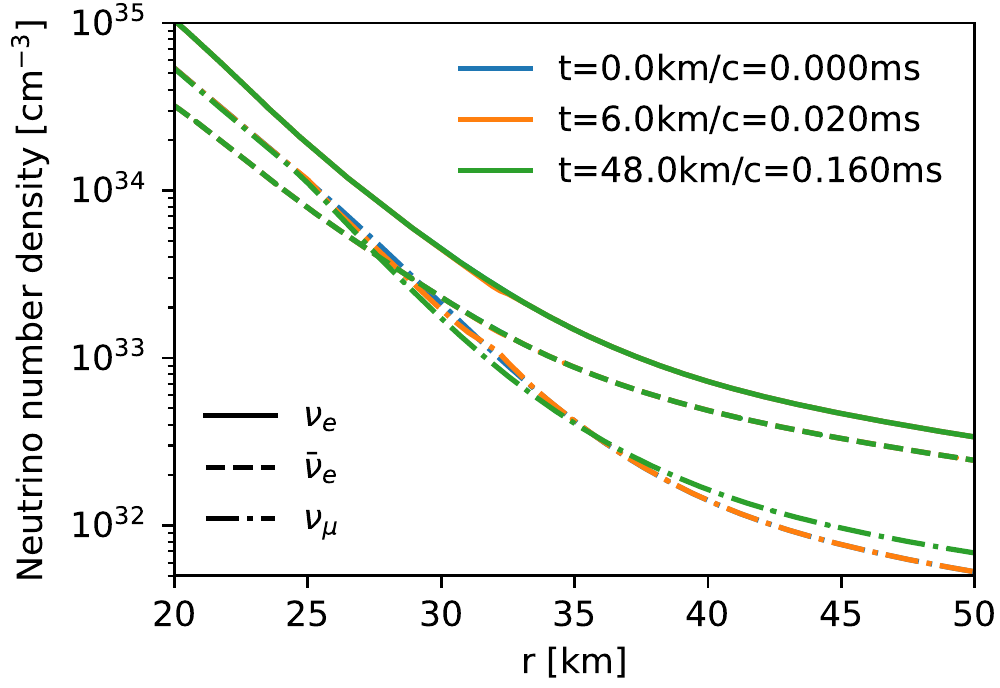}
\llap{\parbox[b]{4.7in}{\small (a)\\\rule{0ex}{1.9in}}} }
\subfigure{\includegraphics[width=0.45\textwidth]{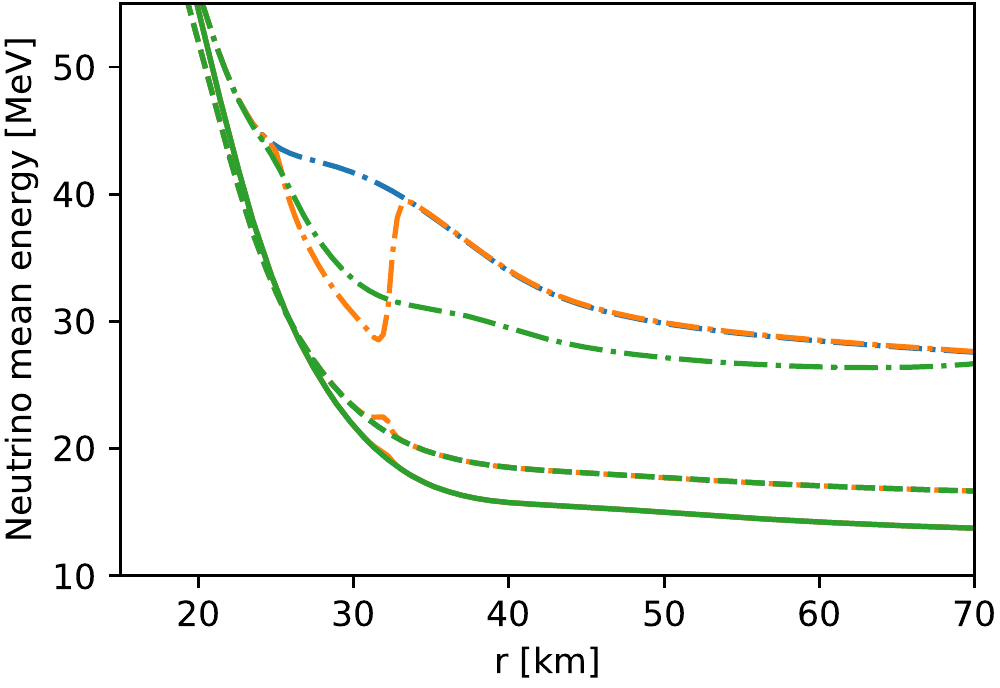}
\llap{\parbox[b]{5.4in}{\small (b)\\\rule{0ex}{1.9in}}} }
\caption{\label{fig:re_II} Radial profiles of neutrino number densities [panel (a)] and mean energies [panel (b)] for $\nu_e$(solid), $\bar\nu_e$ (dashed), and $\nu_\mu$ (dash-dotted) at two radii (32~km and 55~km). Blue, orange, and green colors correspond to three different simulation times 0.0~ms, 0.02~ms, and 0.16~ms respectively.}
\end{figure}

In addition to the energy spectra, we examine two energy-integrated quantities, the neutrino number density and mean energy, to characterize the transport of collision-induced neutrino flavor conversion.
Figure~\ref{fig:re_II} shows the radial profile of these two quantities at different times.
The $\nu_e$ and $\bar\nu_e$ number densities are barely affected except for the tiny dip in $\nu_e$ around 31~km at $t=0.02$~ms.
For $\nu_\mu$, slight changes of $n_{\nu_\mu}$ due to flavor conversion around 32~km are present. More clearly visible is the larger enhancement of $n_{\nu_\mu}$ at $t=0.16$~ms for $r\gtrsim 35$~km due to the extra contribution from the converted ones at the low energy part of the spectrum discussed above.
Although the impact on the number densities seem to be small, flavor conversion and transport due to collisional instability does lead to significant changes to the mean neutrino energies, particularly for $\nu_\mu$.
Figure~\ref{fig:re_II}(b) shows that a large decrease of $\langle E_{\nu_\mu}\rangle$ from $\simeq 40$~MeV locally around $r\simeq 30$~km to as low as $\lesssim 30$~MeV at $0.02$~ms due to flavor conversion.  
Once again, this effect gets transported outward and leaves behind a lowered $\langle E_{\nu_\mu}\rangle$ for $r\gtrsim 25$~km at e.g., $t=0.16$~ms.

\subsection{Models I, III, and IV}
\begin{figure*}
\includegraphics[width=0.99\textwidth]{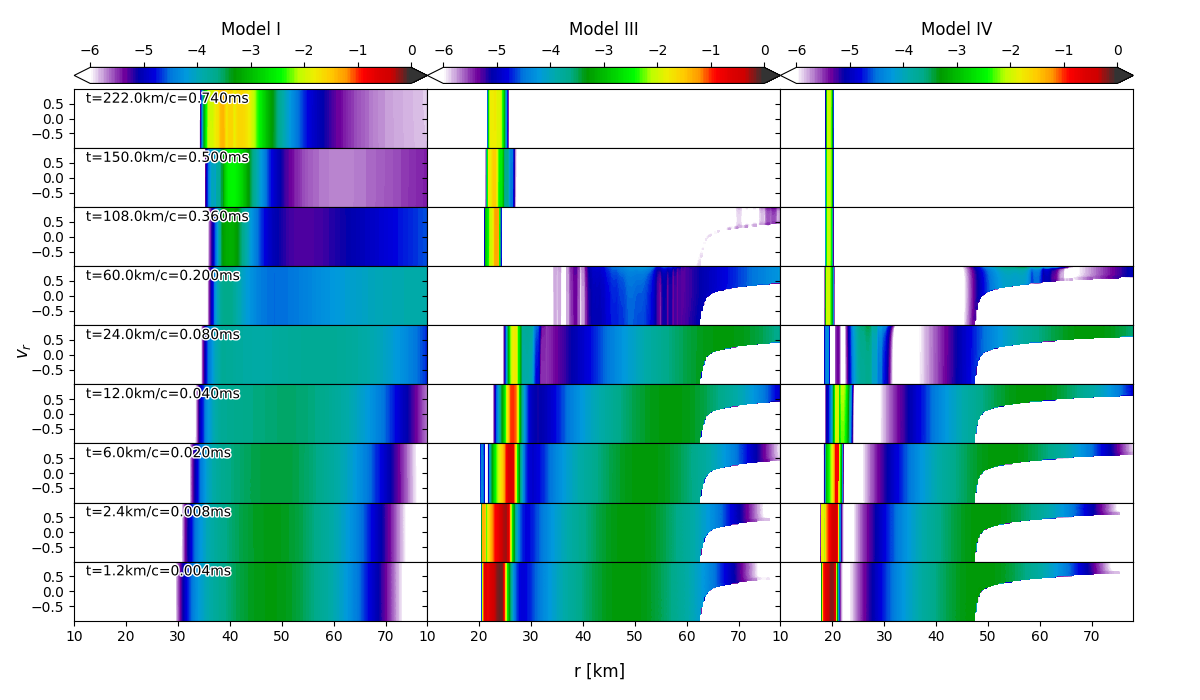}
\caption{\label{fig:2d_III} Evolution of the dimensionless ratio $\log_{10}(s_{e\mu})$ in models I (left panel), III (middle panel) and IV (right panel) at different simulation times. The lower-right corners in the middle and right panels have too small values of $|\langle \varrho_{ee} \rangle_E-\langle \varrho_{\mu\mu} \rangle_E|<10^{29}~{\rm cm}^{-3}$ and hence not shown.}
\end{figure*}

After examining the simulation results in Model II, we now discuss the similarity and differences obtained in Model I, III, and IV, based on different background SN profiles taken from the \texttt{BOLTZTRAN} simulation.
Figure~\ref{fig:2d_III} shows the evolution of $s_{e\mu}$ for these three models. 
For Model I, the collisional instabilities are not strong enough at the beginning so the Gaussian perturbation simply gets transported outward. 
At $t\sim 0.2$~ms, an unstable mode grows around $r\sim 38$~km.
The growth rate is significantly lower than that in Model II since the dimensionless ratio reaches at maximum $\sim 5\times 10^{-3}$ at $t=0.5$~ms and $\sim 5\times 10^{-2}$ at $t=0.74$~ms.

For Models III and IV, the flavor conversion and transport behave similarly to Model II.
Collisional instabilities develop immediately around 20--25~km and 18--22~km in these two models, respectively. 
Afterwards, they grow to nonlinear regime and bifurcate. 
Once again, the inward going mode gets suppressed and the outward going mode sustains and propagate to larger radii.  
Also, second onsets of flavor instability happen in both models at $t\simeq 0.36$~ms and $t\simeq 0.2$~ms and seem to reach stationary states without propagating outwards at the end of our simulations.

\begin{figure*}[t]
\subfigure{\includegraphics[width=0.45\textwidth]{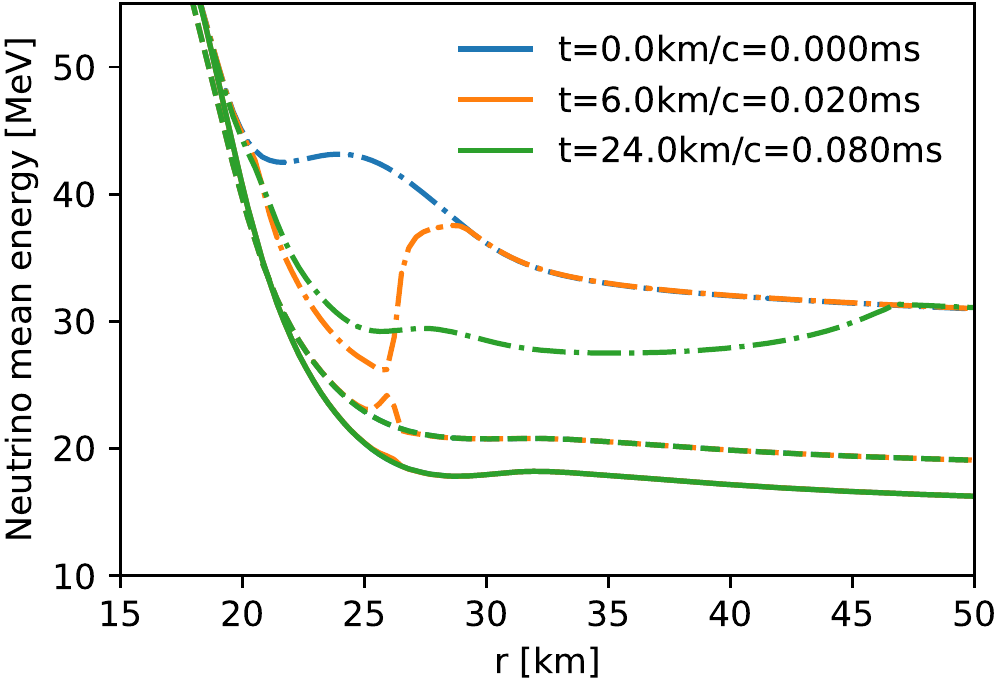}
\llap{\parbox[b]{5.3in}{\small (a)\\\rule{0ex}{1.9in}}} }
\subfigure{\includegraphics[width=0.45\textwidth]{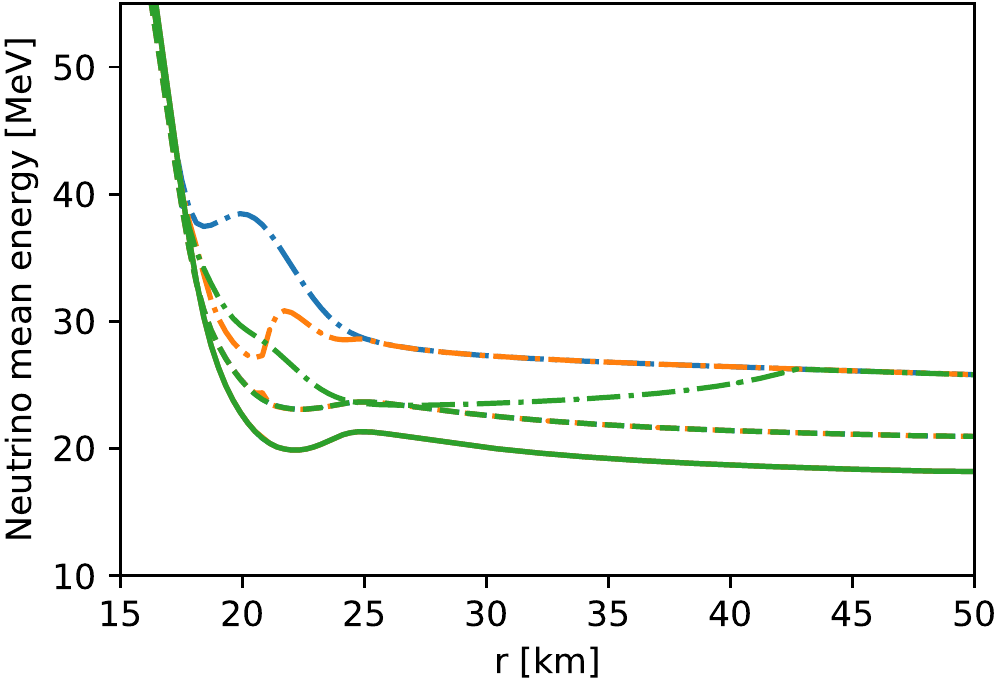}
\llap{\parbox[b]{5.1in}{\small (b)\\\rule{0ex}{1.9in}}} }
\caption{\label{fig:re_III} Radial profiles of neutrino mean energies for $\nu_e$(solid), $\bar\nu_e$ (dashed), and $\nu_\mu$ (dash-dotted) in model III [panel (a)] and model IV [panel (b)]. Blue, orange, and green colors correspond to three different simulation times 0.0~ms, 0.02~ms, and 0.08~ms respectively.}
\end{figure*}

As in Model II, the effect of flavor conversions on radial profile of number densities in all three models are limited and we do not present them explicitly. 
For the neutrino mean energies, we show results from Model III and IV in Fig.~\ref{fig:re_III}.
Similar to Fig.~\ref{fig:re_II}, the collisional instabilities lead to significant reductions of $\langle E_{\nu_\mu}\rangle$ by $\sim\mathcal O(10)$~MeV and slightly increases $\langle E_{\nu_e}\rangle$ and $\langle E_{\bar\nu_e}\rangle$ at where the instabilities occur.
The changes in $\nu_\mu$ are carried to large radii, while strong EA collisions of $\nu_e$ and $\bar\nu_e$ refrains their local changes from propagating outwards.

The general behaviors in those four models can be understood with the simple criteria of asymmetries in neutrino number densities and the collision rates in Eq.~\eqref{eq:LSA_Im}.
Given that $C_{\rm e\mu,EA}/\bar C_{\rm e\mu,EA}  \approx \chi_e/\bar \chi_e \approx (1-Y_e)/Y_e$, the asymmetry factor of EA rate is $\alpha_C\simeq |1-2Y_e|$.
Neglecting the higher order correction, Eq.~\eqref{eq:LSA_Im} predicts a local instability when $\alpha_C>\alpha_n$ with a growth rate proportional to $|C_{\rm e\mu,EA} + \bar C_{\rm e\mu,EA}|\times(\alpha_C/\alpha_n-1)$.

\begin{figure}[t]
\includegraphics[width=0.48\textwidth]{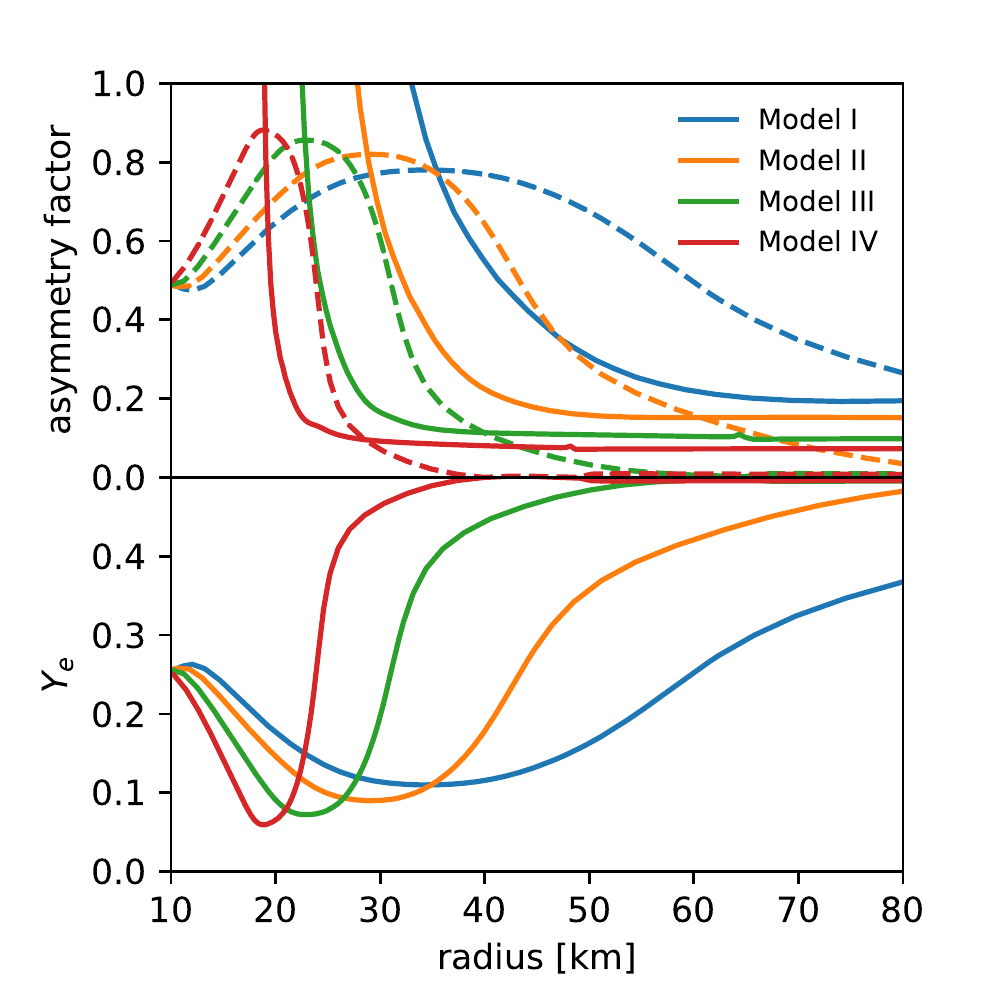}
\caption{\label{fig:criteria} Profiles for the asymmetry factor $\alpha_n$ (solid), $|1-2Y_e|\simeq \alpha_C$ (dashed) in the upper panel, and $Y_e$ in the lower panel at four CCSN snapshots.
An approximate condition for collisional instability is $\alpha_C > \alpha_n$. }
\end{figure}

Figure~\ref{fig:criteria} compares $|1-2Y_e|\simeq \alpha_C$ (dashed curves) to $\alpha_n$ (solid curves) and shows the corresponding $Y_e$ profiles in the bottom panel.
The deleptonization in CCSN leads to more neutron-rich condition near the neutrinosphere at the later time so that the maximal value of $|1-2Y_e|\simeq\alpha_C$ increases with time.
For $\alpha_n$, it is larger than 1 in denser region where $n_{\nu_e}>n_{\nu_\mu}>n_{\bar\nu_e}$ but decreases with radius and asymptotic to a value smaller than 1 for all four snapshots.
More importantly, the asymptotic $\alpha_n<1$ is lower at a later SN snapshot considered.
This is because the SN evolves from the initial stage of neutronization burst dominated by $\nu_e$ emission toward the accretion phase where the difference between $\nu_e$ and $\bar\nu_e$ becomes smaller, thus leading to smaller values of the asymptotic $\alpha_n$.
Consequently, $\alpha_C/\alpha_n>1$ become larger around the neutrinosphere in a later SN profile.
Moreover, the collision rates also increases as the matter density around the neutrinosphere increases over time. 
Combining all these factors described above, it is clear that later SN stages contain more favorable conditions for the collisional instabilities to grow against advection, consistent with our simulation results.

Our results obtained above in Models I--IV tend to suggest that the collisional instability in spherically symmetric SN models can affect the properties of heavy lepton flavor neutrinos at and beyond their decoupling region, but does not lead to major impact on $\nu_e$ and $\bar\nu_e$ spectra above their neutrinospheres.
Thus, the effect of collisional instability are more likely to manifest in the CCSN dynamics and the emitted neutrino signals, and may moderately affect neutrino (induced) nucleosynthetic processes that are sensitive to the spectra of heavy-lepton neutrino flavors.  
For the condition of nucleosynthesis in the neutrino driven wind, the impact of collisional instability may be minor since the outcome depend more on the properties of $\nu_e$ and $\bar\nu_e$.

\section{Effects of other terms on collisional instability}
\label{sec:effects}
\begin{figure*}
\includegraphics[width=0.99\textwidth]{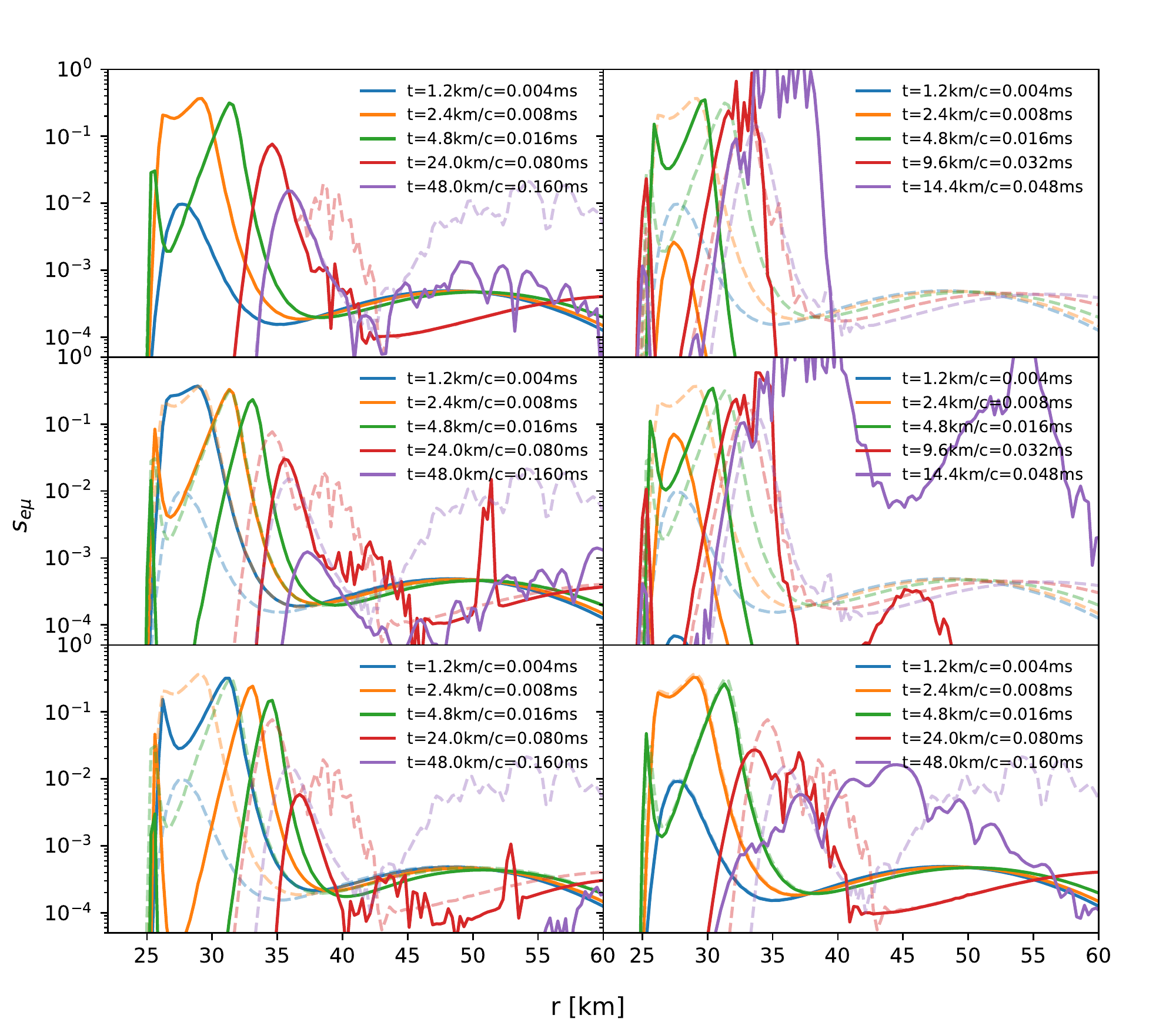}
\llap{\parbox[b]{12.2in}{\small (a) IIn\\\rule{0ex}{5.5in}}}\hspace{-0.03in}
\llap{\parbox[b]{12.2in}{\small (b) IIe1\\\rule{0ex}{3.8in}}}\hspace{-0.03in}
\llap{\parbox[b]{12.2in}{\small (c)  IIe2\\\rule{0ex}{2.1in}}}\hspace{-0.03in}
\llap{\parbox[b]{6.4in}{\small (d) IIv1\\\rule{0ex}{5.5in}}}\hspace{-0.03in}
\llap{\parbox[b]{6.4in}{\small (e) IIv2\\\rule{0ex}{3.8in}}}\hspace{-0.03in}
\llap{\parbox[b]{6.4in}{\small (f) IIm\\\rule{0ex}{2.1in}}}\hspace{-0.03in}
\caption{\label{fig:ro_all} Radial profiles of dimensionless ratio $s_{e\mu}$ of radial velocity $v_r=1$ in exploratory models IIn, IIe1, IIe2, IIv1, IIv2, and IIm [from panel (a) to (f) respectively]. The simulation times are indicated by different colors as listed in the legend. Results from model II for each simulation time are presented by transparent and dashed curves for comparison.}
\end{figure*}

\begin{figure*}
\includegraphics[width=0.99\textwidth]{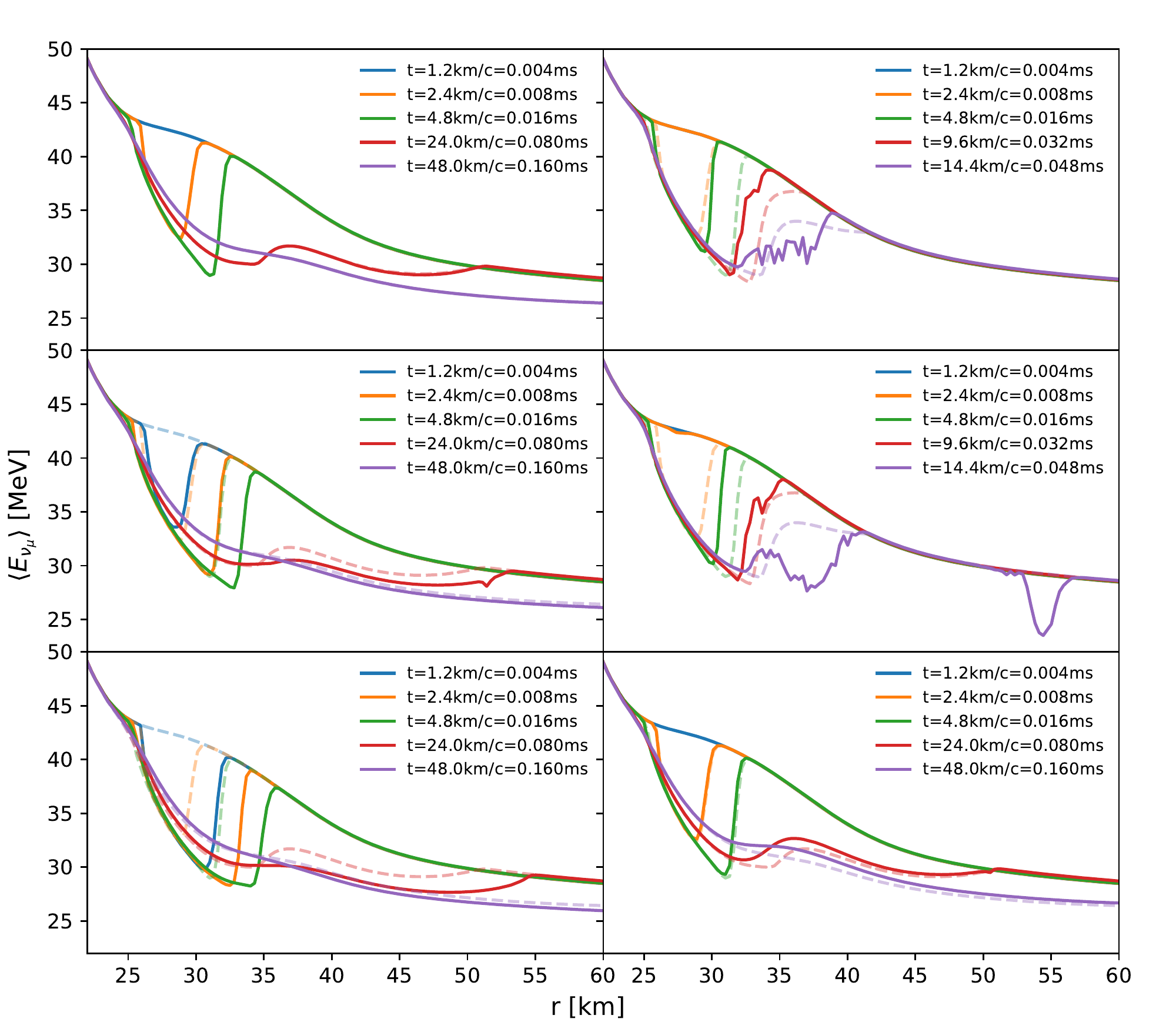}
\llap{\parbox[b]{12.2in}{\small (a) IIn\\\rule{0ex}{5.6in}}}\hspace{-0.03in}
\llap{\parbox[b]{12.2in}{\small (b) IIe1\\\rule{0ex}{3.8in}}}\hspace{-0.03in}
\llap{\parbox[b]{12.2in}{\small (c)  IIe2\\\rule{0ex}{2.in}}}\hspace{-0.03in}
\llap{\parbox[b]{6.2in}{\small (d) IIv1\\\rule{0ex}{5.6in}}}\hspace{-0.03in}
\llap{\parbox[b]{6.2in}{\small (e) IIv2\\\rule{0ex}{3.8in}}}\hspace{-0.03in}
\llap{\parbox[b]{6.2in}{\small (f) IIm\\\rule{0ex}{2.in}}}\hspace{-0.03in}
\caption{\label{fig:re_all} Radial profiles of neutrino mean energies for $\nu_e$(solid), $\bar\nu_e$ (dashed), and $\nu_\mu$ (dash-dotted) in exploratory models IIn, IIe1, IIe2, IIv1, IIv2, and IIm [from panel (a) to (f) respectively]. The simulation times are indicated by different colors as listed in the legend. Results from model II are presented in more transparent colors for each simulation time for comparison.}
\end{figure*}

As mentioned earlier, the typical timescale for collisional instability is in similar order as the inverse of collisional rates. 
This makes this problem different from e.g., the fast flavor conversion for which effects from the vacuum mixing, matter inhomogeneity, and collisions may be ignored to a good approximation for local simulations \cite{martin2020dynamic, bhattacharyya2021fast, bhattacharyya2020late, wu2021collective, zaizen2021nonlinear, abbar2022suppression}. 
In the previous section, we ignored the vacuum term $\mathbf{H}_{\rm vac}$, the matter term $\mathbf{H}_{\rm mat}$, and the impact of the NNS collision on the off-diagonal elements of the neutrino density matrices, for the purpose of purely investigating the outcome of collision instability. 
In this section, we explore consequences of including these terms each by each, as well as effects due to the artificially enhanced decoherence in EA based on the SN background used for Model II.
These six additional models are listed in the second part of Table~\ref{tab:parameters}. The simulation results for all of them are summarized in Figs.~\ref{fig:ro_all} and \ref{fig:re_all}: The former shows the radial profile of $|\langle \varrho_{e\mu} \rangle_E|$ for the angular mode $v_r=1$ at different times, while the latter shows the mean energies of $\nu_\mu$ as a function of radius at the same simulation times. 
Below, we discuss the results of the impact due to the NNS induced decoherence, the artificially enhanced EA decoherence, the matter term, and the vacuum term in each of the subsections.

\subsection{NNS decoherence}
We investigate the decoherence effect due to the NNS by setting the parameter $b_{\rm NNS}=1$ in Model IIn, which allows nonzero off-diagonal elements in $\mathbf C_{\rm NNS}$ and $\bar{\mathbf C}_{\rm NNS}$ that were artificially suppressed in Model I--IV.
Figure~\ref{fig:ro_all}(a) shows that the evolution during the initial phase of $t\lesssim 0.16$~ms is extremely insensitive to the decoherence of NNS as the curves from Model II and Model IIn overlap.
Deviations start to appear at $r\gtrsim 36$~km after $t\gtrsim 0.04$~ms when the flavor mixing propagate outwards. 
The NNS decoherence effect in Model IIn suppresses the amount of flavor mixing by a factor of $\sim 10$ for all velocity modes than the result obtained in Model II.
The main reason that the NNS decoherence effect only shows up at later times at larger radii is related to the angular distribution of neutrinos at locations where flavor mixing occurs. 
At $r\sim 30$~km (inside the neutrinosphere) where initially the instability grows, $\varrho_{e\mu}$ are nearly isotropic (see Fig.~\ref{fig:2d_II}) such that the NNS does not affect their evolution [see Eq.~\eqref{eq:CNNS_chiNNS}].
However, at larger radii where $\varrho_{e\mu}$ becomes anisotropic in $v_r$, the NNS decoherence term not only isotropizes $\varrho_{e\mu}$, but also leads to damping of $\varrho_{e\mu}$ due to the interplay with $H_{\nu\nu}$.
Since the NNS decoherence term mainly affects $\varrho_{e\mu}$, including this term does not lead to any changes to the averaged neutrino energy shown in Fig.~\ref{fig:re_all}(a).

\subsection{Enhanced EA decoherence}
The next models IIe1 and IIe2 assume artificially enhanced off-diagonal elements in $\mathbf C_{\rm EA}$ and $\bar{\mathbf C}_{\rm EA}$ to examine how the results depend on the size of these terms, which are the major source of the collisional instability.
We take the enhancement factors $b_{\rm EA}=2$ in Model IIe1 and $b_{\rm EA}=4$ in Model IIe2.
Panel~(b) and (c) in Fig.~\ref{fig:ro_all} show that for most radii, $s_{e\mu}$ at $0.004$~ms in Models IIe1 and IIe2 are similar to the values at $0.008$~ms and 0.016~ms in Model II, respectively.
This is consistent with the conclusion obtained with the LSA in Sec.~\ref{sec:linear}, because only the off-diagonal elements of $C_{e\mu,\rm EA}$ and $\bar C_{e\mu,\rm EA}$ enter the linearized equation.

After the instability grows to the linear regime and propagate outwards, the evolution of flavor mixing are also clearly affected by the enhancement. 
Overall, the mixing propagates faster with larger values of $b_{\rm EA}$ but also gets damped earlier (see e.g., the curves at 0.08 and 0.16~ms). 
The effect of faster propagation is also clearly shown in the evolution of $\langle E_{\nu_\mu}\rangle$ shown in panels (b) and (c) of Fig.~\ref{fig:re_all}.

\subsection{Vacuum term}\label{sec:vacuumterm}
Models IIv1 and IIv2 include the vacuum Hamiltonian ${\mathbf H}_{\rm vac}$. 
In both models, we assume an effectively reduced mixing $\theta_V=10^{-6}$. 
For $\delta m^2$, we take different values of $\delta m^2=\delta m^2_{\odot}=8\times 10^{-5}~{\rm eV}^2$ for Model IIv1 and $\delta m^2=\delta m^2_{\rm atm}=2.3\times 10^{-3}~{\rm eV}^2$ for Model IIv2, which correspond to the measured values in solar and atmospheric neutrino experiments,  respectively.
For these models, we do not apply any initial perturbation in the off-diagonal terms of $\varrho$ and $\bar\varrho$ as flavor mixing is directly generated by ${\mathbf H}_{\rm vac}$.

Panels~(d) and (e) of Fig.~\ref{fig:ro_all} show that the collisional instabilities develop in both models and reach the nonlinear regime within a similar timescale $\sim \mathcal O(0.01~{\rm ms})$, similar to the case using artificial initial Gaussian perturbation.
Because Model IIv2 has larger $\delta m^2$ than Model IIv1, which generates larger flavor mixing seed,
$s_{e\mu}$ grows faster in Model IIv2 in the linear regime.

For both models, we find that the inclusion of $H_{\rm vac}$ tends to help amplify flavor mixing when they propagate outwards, as shown by the larger value of $s_{e\mu}$ around $r\simeq 38$~km at $t=0.048$~ms (purple line). 
As a result, the corresponding $\langle E_{\nu_\mu}\rangle$ in these models also become smaller; see panel (d) and (e) in Fig.~\ref{fig:re_all}.

In Model IIv2 with a larger $\delta m^2$, another instability occurs at $r\sim 45$~km at 0.032~ms, which later grows to the nonlinear regime. 
This flavor conversion is not triggered by the collisional instability, but due to the well studied slow mode (see Sec.~\ref{sec:introduction}).
However, we note that this onset of slow mode in our simulation domain is related to the fact that we have attenuated $\mathbf{H}_{\nu\nu}$ and  neglected $\mathbf{H}_{\rm mat}$. 
Unlike the collisional instability which is not sensitive to these assumptions, the slow mode instability can be suppressed by large values of $V_{\nu\nu}$ and $V_{\rm mat}$~\cite{Esteban-Pretel:2008ovd,Duan:2010bf,Chakraborty:2011nf}, and should only appear at much larger radii if we had used $V_{\nu\nu}$ and $V_{\rm mat}$ from the corresponding SN simulation snapshot.

\subsection{Matter term}\label{sec:matter}
An ingredient commonly ignored in the flavor evolution equation of homogeneous oscillation models is the matter term.
In those models, the medium density is assumed to be homogeneous such that it effectively introduces a global phase to all considered neutrinos, which can be rotated away~\cite{Duan:2005cp}.
However, since the length scale of collisional instability is $\sim\mathcal O$(1)~km, comparable to the scale length of varying matter profile, the contribution of ${\mathbf H}_{\rm mat}$ at different locations varies, which may modify the dispersion relation of the flavor wave.

Since the original $V_{\rm mat}$ from SN simulation snapshot has the largest values in our simulation domain (see Fig.~\ref{fig:rates_comparison}), whose corresponding length scale of oscillations cannot be resolved by our simulation setup, we take a parametric function
\begin{equation}
V_{\rm mat} = (2\Delta r)^{-1} \exp\left[ -(r-10~{\rm km})^4/(18~{\rm km})^4 \right]
\end{equation}
shown by the blue dotted curve in Fig.~\ref{fig:rates_comparison} to probe the effect of inhomogeneous matter profile on the evolution of collisional instability.
This parametric function is taken in a way that it mainly varies between 20~km and 50~km where the collisional instability occurs.
It becomes a constant when $r<20$~km and vanishes to zero when $r>50$~km.

Figure~\ref{fig:ro_all}(f) shows that although this inhomogeneous matter term does not affect the initial evolution of flavor instability for $t\leq 0.016$~ms, it does affect the later transport of flavor mixing by mainly reducing the group velocity of the propagating flavor waves than that obtained in Model II without any matter term, which is clearly demonstrated by the red and purple curves at $t=0.08$ and $0.16$~ms shown in the same panel.
Once again, this effect is also reflected in $\langle E_{\nu_\mu}\rangle$ shown in Fig.~\ref{fig:re_all}(f), where the reduction of the average energy spreads to larger radii more gradually compared to Model II.

\section{Discussion and conclusions}
\label{sec:discussion}
We have implemented a multi-group and discrete-ordinate collective neutrino oscillation simulator, and solved the neutrino QKE in a spherically symmetric geometry, including global advection as well as realistic collisional rates in a self-consistent way.
We used this simulator to study the occurrence and transport of collisional instability in the absence of fast flavor conversion.  
Our simulations were performed within the radial range of 10~km to 85~km under hydrostatic backgrounds taken at different stages from a CCSN simulation.
We confirmed the existence of collisional instability near the neutrinosphere, and found that the strength of instability increases in later SN stages which have more asymmetric EA rates due to the deleptonization of the matter and less differences between $\nu_e$ and $\bar\nu_e$ number densities.

We found that the collisional instability can lead to significant flavor conversions in three fiducial models that we examined (Models II-IV), which included minimal ingredients to trigger the instability.
Flavor mixing developed at the location of the initial instability can be transported both inwards and outwards.
The inner branch gets damped and reset by collisions, while the outer branch transports to the free-streaming regime.
For electron (anti)neutrinos, although their distributions are affected at the onset of the flavor conversion, the relatively large EA rates quickly reset their spectra close to the equilibrium ones.
For heavy lepton (anti)neutrinos, flavor conversions not only affect their distributions near their decoupling region, but also leave imprints in their spectra at the free-streaming regime.
Overall, their neutrino number densities are slightly increased and their mean energy are significantly reduced.
In one of the four models (Model I), we found that the growth rate of the flavor instability is too small against advection so that flavor mixing does not reach the nonlinear regime at the end of our simulation time $\sim 1$~ms.

The results derived in this work suggest a major difference from Ref.~\cite{johns2021collisional}: Although (anti)neutrinos of heavy lepton flavors are affected, their number densities do not converge to those of electron (anti)neutrinos as predicted in the homogeneous model wherein global advection is absent and therefore flavor conversion runs to completion. 
Our results also imply that although the collisional instability may not directly affect the $\nu_e$ and $\bar\nu_e$ heating rates behind the accretion shock, the altered emission of heavy lepton flavors from region around neutrinosphere may still have potential impact on supernova dynamics. 
In addition, the changes of the energy spectra of heavy lepton flavors can be probed by the future detection of nearby CCSNe with high statistics. 

Beyond the fiducial models, we have also examined impacts from different terms in the QKE, including the effects due to the NNS decoherence, artificially enhanced EA decoherence, neutrino vacuum mixing, and inhomogeneous matter profile for Model II.
Including the off-diagonal element of the NNS collisional term introduces little changes to inner regions where neutrino distributions are nearly isotropic, but damps flavor mixing at larger radii.
The artificially enhanced EA decoherence leads to higher growth rates of flavor instability.
With non-zero vacuum mixing, the flavor instability reaches the nonlinear regime earlier.
It also results in further decreased mean energy of heavy lepton flavor neutrinos.
The inhomogeneous matter potential mainly changes the group velocity of the flavor mixing wave by alternating the dispersion relation.
We note that although each of these terms affects the quantitative behavior of the system, the qualitative features demonstrated by our fiducial models remain robust.

Although our results confirmed and provided important insights to understand the collisional flavor instability in SN environment, some cautions should be noted.
First, our simulations are restricted by the trilemma among self-consistency, advection, and exact rates and hence rely on the strategical attenuation and parametrization of coherence scattering potentials that may give rise to inaccurate results when the vacuum mixing and matter potential are included. 
Second, due to the computational limitation, we did not include the NES contribution to collisions, which are important in determining the exact mean energy of heavy lepton flavors despite their subdominant role in trapping them. 
It will be crucial to include these rates in future to assess the actual impact of collisional flavor instabilities on heavy lepton flavors. 
Moreover, our models are based on static SN backgrounds and assume spherical symmetry, thus neglects the potential impacts due to dynamic evolution of background profiles and the anisotropy, as well as the feedback of flavor instability on SN evolution.  
It remains to be seen how the conclusion derived in this work holds in a more complete SN model where these simplifications and assumptions are addressed. 

\begin{acknowledgments}
We thank Gang Guo and Ninoy Rahman for discussions on the neutrino collisional rates.
ZX and GMP acknowledge support by the European Research Council (ERC) under the European Union’s Horizon 2020 research and innovation programme (ERC Advanced Grant KILONOVA No.~885281), the Deutsche Forschungsgemeinschaft (DFG, German Research Foundation) -- Project ID No.~279384907 - SFB 1245 ``Nuclei: From Fundamental Interactions to Structure and Stars'', and the State of Hesse within the Cluster Project ELEMENTS.
MRW and MG acknowledge supports from the National Science and Techonology Council, Taiwan under Grant No.~110-2112-M-001-050, No.~111-2628-M-001-003-MY4, and the Academia Sinica (Project No.~AS-CDA-109-M11).  
MRW also acknowledges supports from the Physics Division of the National Center for Theoretical Sciences, Taiwan.
TF acknowledges support from the Polish National Science Center (NCN) under Grant No.~2020/37/B/ST9/00691.
LJ was supported by NASA through the NASA Hubble Fellowship Grant No.~HST-HF2-51461.001-A awarded by the Space Telescope Science Institute, which is operated by the Association of Universities for Research in Astronomy, Incorporated, under NASA Contract No.~NAS5-26555.
We would like to acknowledge the use of the following software: \texttt{Matplotlib} \cite{matplotlib}, \texttt{Numpy} \cite{numpy}, and \texttt{SciPy} \cite{scipy}.
\end{acknowledgments}

\bibliographystyle{apsrev4-1}
\bibliography{references.bib}

\appendix
\section{Rates of collisions}
\label{sec:A1}

The emissivity and opacity for EA processes are taken from \texttt{BOLTZTRAN} using relativistic dispersion relations for nucleons in the nuclear medium and including weak magnetism, pseudoscalar term, and form factor effects \cite{guo2020charged}.

NNS is almost elastic and commonly treated as iso-energetic so that
\begin{align}
    & \mathbf C_{\rm NNS}(E, v_r) = \frac{1}{2} \int d v_r'\, \times \nonumber\\
    & \left[ - R_{\rm NNS}(E, v_r, v_r') \left\{ \varrho(E, v_r), \mathcal I- \frac{4\pi^2}{E^2} \varrho(E, v_r') \right\} \right. \nonumber\\
    & \left. + R_{\rm NNS}(E, v_r', v_r) \left\{ \varrho(E, v_r'), \mathcal I- \frac{4\pi^2}{E^2} \varrho(E, v_r) \right\} \right],
\end{align}
where, e.g., $4\pi^2 \varrho_{ee}(E, v_r)/E^2 $ inside the anti-commutator is the distribution function of $\nu_e$ accounting for the Pauli blocking.
This iso-energetic scattering kernel has the symmetry $R_{\rm NNS}(E, v_r, v_r')=R_{\rm NNS}(E, v_r', v_r)$ so that the blocking in the gain and loss terms cancels as $\{\varrho(E,v_r),\varrho(E,v_r')\} = \{\varrho(E,v_r'),\varrho(E,v_r)\}$ and effectively has no impacts.

The two moments of NNS opacities in Eq.~\eqref{eq:CNNS_chiNNS} without weak-magnetism and recoil correction are given as below
\begin{align}
	\chi_{\rm NNS} (E) & = \frac{G_F^2 E^2}{\pi} [(c_{V,n}^2+3 c_{A,n}^2)\eta_n + (c_{V,p}^2+3 c_{A,p}^2)\eta_p], \nonumber\\
    \tilde\chi_{\rm NNS} (E) & = \frac{G_F^2 E^2}{\pi} [(c_{V,n}^2- c_{A,n}^2)\eta_n + (c_{V,p}^2- c_{A,p}^2)\eta_p],
\end{align}
with weak coupling constants $c_{V,p} = 1/2-2\sin^2 \theta_{\rm W}$, $c_{V,n}=-1/2$, $c_{A,p} = g_A/2$, $c_{A,n}=-g_A/2$, $g_A\approx 1.27$, and $\sin^2 \theta_{\rm W}\approx 0.23$.
The effective number densities
\begin{equation}
	\eta_N = \frac{3Tn_N}{\sqrt{4E_{{\rm F}, N}^2+9T^2}}
\end{equation}
account for the nucleon final state blocking with Fermi energies $E_{{\rm F}, N}=(3\pi^2 n_N)^{2/3}/(2 m_u)$, and number densities $n_N = \rho Y_N/m_u$ for $N=n,p,$ respectively.

\section{Resolution}
\label{sec:A2}
\begin{figure*}[t]
\includegraphics[width=0.99\textwidth]{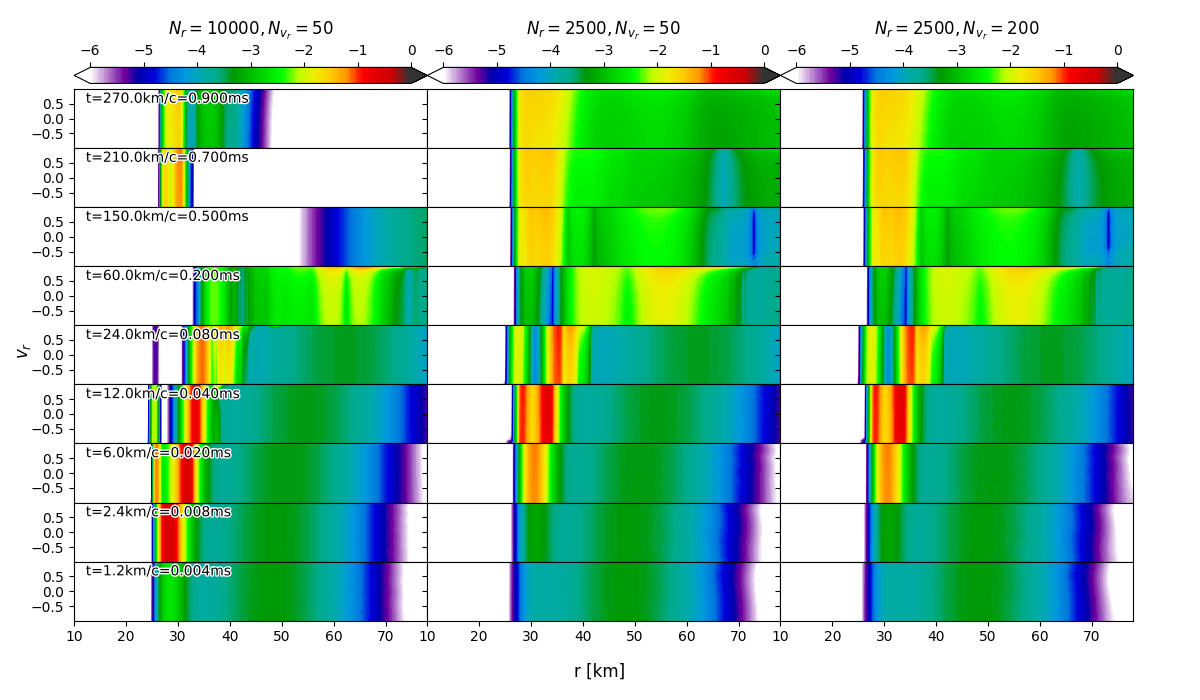}
\caption{\label{fig:2d_resolution} Evolution of the dimensionless ratio $\log_{10}(s_{e\mu})$ in Model II at $t_{\rm pb}=247$~ms with three resolutions: $N_r=10000, N_{v_r}=50$ (left panel), $N_r=2500, N_{v_r}=50$ (middle panel), and $N_r=2500, N_{v_r}=200$ (right panel).}
\end{figure*}

In this section, we discuss the dependence of our results on the adopted simulation resolutions for Model II.
In addition to the fiducial models with highest radial resolution $N_r=25000$, we examine the other two different radial resolutions with $N_r=10000$ and 2500.
For models with lower radial resolutions, we use the same parameters for Model II listed in Table~\ref{tab:parameters}, except $a_1=4\times 10^{-4}$ for $N_r=10000$ and $a_1=10^{-4}$ for $N_r=2500$. 
The different choice of $a_1$ here is to satisfy the requirement that the length scale of oscillations can be resolved by $\Delta r$.
We have also examined cases with $N_{v_r}=200$ for the lowest radial resolution runs.
The evolution histories are shown in Fig.~\ref{fig:2d_resolution}.
The case with $N_{v_r}=200$ undergoes almost identical evolution as the one with $N_{v_r}=50$ which shows that 50 angular grids are sufficient.

With the same number of angular grids $N_{v_r}=50$, the lowest radial resolution run ($N_r=2500$) shows clearly visible differences from results obtained with intermediate ($N_r=10000$) and high ($N_r=25000$) resolutions.
First of all, the grow rate in the initial linear regime is reduced while the onset radius of instability shifts from $r\approx 29$~km to $32$~km.
Similar to runs with higher resolutions, after the flavor mixing reaches nonlinear regime between $t= 0.02$ and 0.04~ms, it bifurcates. 
However, the inward-moving branch does not get damped completely, regrows around $t \sim 0.2$~ms, and eventually reaches the stationary state much earlier than the high resolution runs by $\approx 0.5$~ms.
On the other hand, results obtained with $N_r=10000$ and $N_r=25000$ (see Fig.~\ref{fig:2d_II}) are qualitatively more similar, although slight differences exist.

\begin{figure}[t]
\includegraphics[width=0.45\textwidth]{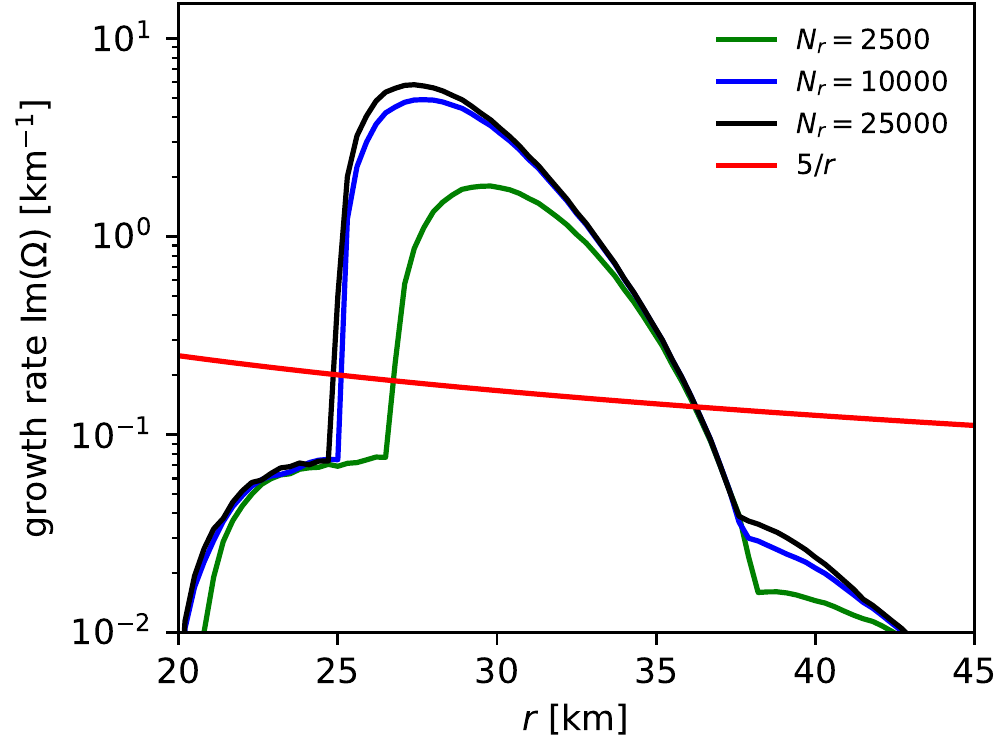}
\caption{\label{fig:LSA2} Growth rates Im$(\Omega)$ from linear stability analysis as functions of radius $r$ at $t=0$~ms for three choices of radial resolutions and attenuation factors in Model II. We sample different $K_r$ and show the maximal values of growth rates as in Fig.~\ref{fig:LSA}.
}
\end{figure}

The smaller growth rates in the initial linear regime with $N_r=2500$ is related to the imposed larger attenuation on $\mathbf H_{\nu\nu}$.
As indicated from Eq.~\eqref{eq:LSA_Im}, if $V_{\nu\nu}$ and collisional rates are in similar magnitude, the high-order term can reduces the growth rate of the instability.
On the other hand, if $V_{\nu\nu}$ is much larger than the collisional rates, the growth rate of collisional instability converges to an asymptotic value. 
The LSA analysis results obtained by solving Eqs.~\ref{eq:LEQ1} and \ref{eq:LEQ2} also support this conclusion.
Figure~\ref{fig:LSA2} shows the growth rates as functions of radius for all three radial resolution runs at $t=0$.
The peak of growth rates with $N_r=2500$ has Im$(\Omega)\approx 1.6$~km$^{-1}$ at $\sim 30$~km, lower than other two cases with peak values $\approx 5-6$~km$^{-1}$ at $\sim 26-27$~km.

For the resolution in the energy grid, in addition to $N_E=20$, we have also used $N_E=8$. 
However, we find that in this case each grid has too wide width to accurately model the sharp dependence of the muonic EA rates at $\sim 65$~MeV. 
This insufficient resolution then leads to the artificial amplification of number density $n_{\nu_\mu}$ near the decoupling region and affects our simulation results.
\end{document}